# On Ambiguity and the Expressive Function of Law: The Role of Pragmatics in Smart Legal Ecosystems

Pompeu Casanovas

Artificial Intelligence Research Institute, Spanish National Research Council (IIIA-CSIC)
pompeu.casanovas@iiia.csic.es
Institute of Law and Technology, Universitat Autònoma de Barcelona (IDT-UAB)
pompeu.casanovas@uab.cat
La Trobe LawTech Research Group, La Trobe University, Bundoora, VIC, Australia
p.casanovasromeu@latrobe.aus.edu

**Abstract:** This is a long paper, an essay, on ambiguity, pragmatics, legal ecosystems, and the expressive function of law. It is divided into two parts and fifteen sections. The first part (*Pragmatics*) addresses ambiguity from the perspective of linguistic and cognitive pragmatics in the legal field. The second part (*Computing*) deals with this issue from the point of view of human-centered design and artificial intelligence, specifically focusing on the notion and modelling of rules and what it means to comply with the rules. This is necessary for the scaffolding of *smart legal ecosystems* (SLE). I will develop this subject with the example of the architecture, information flows, and smart ecosystem of OPTIMAI, an EU project of Industry 4.0 for zero-defect manufacturing.

## I First part: Pragmatics

### 1. Presentation

This essay delves into a pivotal matter whose significance has recently dawned upon me, i.e., the interplay between linguistic pragmatics and what is commonly referred to as *the expressive conception of law*, legal norms, or their functions.[1]

---

[1] The *recursive* conception of law defended by ALCHOURRON AND BULYGYN (1991) opposed the *hylemorphic* conception of rules to the *expressive* conception, considering them as conceptual entities independent of the language in which they can be expressed. The discussion is quite interesting because it actually attends to the conception of law that is being defended. We can't go into it here. Suffice it to note that my notion presupposes that law (and systems of regulation) depend on behaviour, including linguistic behaviour. Another use currently occurs in the field of metaethics where it is discussed whether or not expressivism is part of cognitivist positions. That's a discussion I won't get into either. For a general description, cf. CHRISMAN AND MILLER (2014).

This essay is structured into two distinct parts. In the initial one (1-7), I will explore the nexus between ambiguity and law through the lens of linguistic and cognitive pragmatics. Subsequently (8-15), I will pivot towards examining legal ecosystems from the perspective of legal theory and the sciences of design —artificial intelligence and computer science. I will therefore focus on *pragmatic (contextual) ambiguity* in regulatory models. I will also mention the logical formulations, when necessary, but not deepening into it.[2]

I have organized my discussion into a series of coordinated language activities. The progression for the first part entails: 2. *Express*. 3. *Speak*. 4. *Manage*. 5. *Regulate*. 6. *Interpret*. 7. *Signify*. Conversely, the sequence for the second section comprises: 8. *Elucidate*. 9. *Disambiguate*. 10. *Generate*. 11. *Regenerate*. 12. *Comply*. 13. *Execute*. 14. *Contextualize*. Notably, the deliberate conclusion of the initial part with an examination of layers of meaning in law (7), and the subsequent discussion on automatic and/or semi-automatic legal compliance (13-14) in the second part, underscores a strategic narrative. In the closing stages (Section 15. *Re-express*), I will establish the connection between these discussions and advance several theses on ambiguity.

## 2. Express

The conventional understanding of the expressive function of law often pertains to the declarations articulated in the dispositive section of rules, which are not explicitly intended to dictate behaviour. However, I find this perspective somewhat limiting. Both the substance and the enactment of regulation possess an expressive dimension that resonates socially, irrespective of whether they directly influence rights and obligations. In essence, the mere act of regulation or standardization can nullify the intended effects of the regulation. To echo the sentiments of Cass Sunstein (1996: p. 2023):

> Few people have burned the American flag in recent years, and it is reasonable to suppose that a constitutional amendment making it possible to criminalize flag burning would have among its principal consequences a dramatic increase in annual acts of flag burning. *In fact, adopting a constitutional amendment may be the best possible way to promote the incidence of flag burning*. In these circumstances it seems clear that those who support the amendment are motivated not so much by consequences as by *expressive concerns*.

---

[2] I will leave aside the ambiguity in logic. Ambiguity is related to (i) *uncertainty*, (ii) *gradation of truth*, (iii) *different values* to T/F, (iv) and different notions of *consequence* that depart from classic logics, i.e. the so-called *paraconsistent logics* (which reason on inconsistent premises without presupposing the 'explosion' that anything is derived from a contradiction), *preservationism* (the approach to the premises in such a way that the consequence preserves some truth) and *dialetheism* (according to which some contradictions can be true). It should be specified that point (ii) is a particular case of point (iii), if the truth is graded, it needs more values than T/F. Also, that, in point (iii), for example in the so-called substructural logics, the notion of consequence is different from the classical one because they lack some of the structural rules of classical logic (hence its name), and on the other hand, they are not paraconsistent or preservationists or dialethists (I thank LLUÍS GODO for these precisions). On the formalization of uncertainty and the gradation of truth from fuzzy sets, see DUBOIS *et al*. (2007): "fuzzy sets membership grades can be interpreted in various ways which play a role in human reasoning, such as levels of intensity, similarity degrees, levels of uncertainty, and degrees of preference" *(ibid*.: p. 325). BROWN (2007: p. 117) directly connects the notion of consequence with ambiguity: "by treating certain sets of atomic sentences as ambiguous, we can project consistent images of inconsistent premise sets: (i) Γ' is a consistent image of Γ based on A iff (ii) A is a set of sentence letters, (iii) Γ' is consistent, (iv) Γ' results from the substitution, for each occurrence of each member a of A in Γ, of one of a pair of new sentence letters, af and at". This notion of consequence would be identical to that of PRIEST'S logic of paradox (1979, 1984), considered by many as the origins of *dialethical* positions. At the beginning of his article, PRIEST (1979: p. 219) sets out his intuition: "to suggest a new way of handling the logical paradoxes. Instead of trying to dissolve them, or explain what has gone wrong, we should accept them and learn to come to live with them". PRIEST (1984: p.154) extends it to polysemy and homonymy ("multicriterial terms"): "That there are true contradictions will seem to many a radical proposal. However, those reasonably familiar with the history of philosophy will know that it is not a novel one! The logical paradoxes are not the only considerations which can be adduced in favor of this position. For example, analyzes of motion, of inconsistent legal corpuses, and of multicriteria terms have also been known to issue in this conclusion." See also BROWN (2007), and on paraconsistent logics, Priest (2007).

This observation extends beyond specific domains to encompass all socially sensitive issues, such as racism, hate speech, or gender crimes. Social sensitivity is primarily, though not exclusively, addressed through legal frameworks. However, while laws are undoubtedly necessary, the process of legal regulation can pose significant barriers to both the efficacy of the law itself and the desired social outcomes. Unintended consequences and collateral, counter-intentional, or preter-intentional effects are perennial subjects in sociology and criminal law. Furthermore, in the realm of pragmatics, the concept of *pragmatic* and *praxeological* counter-performative verbs, as introduced by ÉMILE BENVENISTE and notably expanded upon by MARIA ELISABETH CONTE[3], sheds light on these phenomena. The efficacy of regulatory mechanisms can be undermined precisely due to their improper application.

In digital realms and contexts, the phenomenon of "illocutionary suicide," to borrow ZENO VENDLER's dramatic terminology[4], is amplified. This is compounded by semantic reflexivity, giving rise to the complexities of digital legal governance that we are just beginning to delineate and explore (CASANOVAS AND NORIEGA, 2022). These complexities intertwine with well-known deontic paradoxes, reminiscent of the classic dilemma —"Either slip the letter into the letter-box or burn it!".[5] The oversight of classical pragmatics in the realm of Artificial Intelligence and law has led to an oversimplified schematization of normative models, impeding their effective implementation in social contexts, including the development of intelligent legal ecosystems, as we will delve into later. For now, it is paramount to acknowledge that deontic modalities, whether verbal or logical, cannot singularly address the spectrum of regulation. Numerous other modes of regulation exist, which manifest not through explicit obligations, prohibitions, or permissions, but rather through the interactive behaviour of agents.

---

[3] Cf. M.E. CONTE (1983, 1984: p. 71-72). "3.3 Je vais faire une distinction entre deux formes de contreperformativité, dont les cas paradigmatiques sont les verbes 'insinuer' et 'to allege'. Le premier cas paradigmatique est 'insinuer'. Si quelqu'un dit : 'J'insinue que *p*', il fait échouer l'acte qu'il dit d'accomplir. L'insinuation échoue justement à cause du fait qu'il a été qualifié d'insinuation. Puisque le moyen exclut le but (l'objectif de l'action), la contreperformativité de 'insinuer' est une contreperformativité *praxéologique* […]. Tandis que pour les verbes performatifs comme 'affirmer' l'usage du verbe à la première personne du présent de l'indicatif es une condition *suffisante* pour l'*accomplissement* de l'acte désigné par le verbe, pour les verbes contreperformatifs comme 'insinuer' l'usage du verbe à la première personne du présent de l'indicatif est une condition *suffisante* pour le *non-accomplissement* de l'acte désigné par le verbe. En disant 'I allege…', le locuteur déclare la non-existence d'une présupposition pragmatique (l'engagement à la vérité de la proposition affirmée) de l'acte virtuel corrélatif à l'*allégation*, et précisément de l'acte linguistique d'assertion. Le posé de 'I allege…' est qu'un *présupposé* de l'assertion n'existe pas. La contreperformativité de 'to allege', puisqu'elle est en relation avec un présupposé pragmatique d'un acte de langage, est une *contreperformativité pragmatique*. 3.4 J'ai donc fait la différence entre une contreperformativité *praxéologique* et une contreperformativité *pragmatique.* Dans le cas de la contreperformativité *praxéologique*, l'énonciation rend vain l'acte même que le locuteur dit d'accomplir. Dans le cas de la contreperformativité *pragmatique*, l'énonciation rend vain l'acte linguistique virtuel corrélatif. ".
[4] As cited by M.E. CONTE, op. cit. ibid. p. 72.

Another perspective on this issue suggests that linguistic indeterminacy, particularly ambiguity, along with vagueness, *inherently* constitute components of regulatory mechanisms. When applied in real-world scenarios, these elements exhibit distinct characteristics compared to when they are articulated in abstract terms. Regulatory and normative systems necessitate what SUNSTEIN (1994a: p. 139) has termed "incompletely theorized agreements on particular cases."[6] Legal practitioners, whether judges or lawyers, "tend to shift to a more specific level when disagreements arise in abstract discussions" (SUNSTEIN 1994b: p. 1736). Implementing abstract principles not only calls for interpretation but also demands a level of specificity achievable only in concrete cases, at a micro level. Additionally, I propose, and this forms the crux of my argument, that it requires a *meso-level*, an intermediary level of abstraction that facilitates both inward and outward perspectives, enabling the requisite linguistic and contextual interpretations. We have coined this approach, the *middle-out and inside-out approach* (CASANOVAS, DE KOKER, AND HASHMI, 2022), which we find particularly beneficial in delineating the functional prerequisites for platforms, web services, and information-processing modules.

As I will try to show: (i) the interpretation of the elements or components of a language (terms, expressions, uses...) is not sufficient for the regulation of the behaviour of its speakers, whether natural or artificial agents; (ii) this behaviour occurs in dynamic changing environments from which emerge collective properties that transcend their context; (iii) we cannot in the sciences of design anticipate all the situations that will give rise to a sustainable ecosystem; (iv) on the other hand, we can instead stipulate the conditions and direction that this emergency must fulfill in order to develop. We deem this perspective relevant for the field known as *Human-Machine Interaction*' (HMI).

---

[5] This is the example of the so-called "Ross' paradox" about imperatives: (i) for the logic of satisfaction: "If the imperative, 'Slip the letter into the letter-box!' is satisfied, then the imperative, 'Either slip the letter into the letter-box or burn it!' is also satisfied"; (ii) for the logic of validity: "If the imperative, 'Slip the letter into the letter-box!' is valid, then the compound imperative, 'Either the letter is to be slipped into the letter-box, or it is to be burnt" is valid too." Cf. ROSS (1944: p. 41).

[6] "How is law possible in a heterogeneous society, composed of people who sharply disagree about basic values? Such disagreements involve the most important issues of social life: the distribution of wealth, the role of race and gender, the nature of free speech and private property. Much of the answer to this puzzle lies in an appreciation of how people who disagree on fundamental issues can achieve incompletely theorized agreements on particular cases." SUNSTEIN (1994a).

## 3. Speak

This transcends mere linguistic disambiguation. It delves into the art of selecting the right instruments to gauge the expressiveness of our words and, crucially, our intentions behind them. Consider the question, 'Do I help you wash the dishes?'—it could be interpreted as offensive in certain couple relationship dynamics today, whereas thirty years ago, it might not have been. [7] Its meaning hinges not just on the words used but on who says them, how, and in what context.

Comprehending the meaning of an expression entails more than just parsing its verbal components; it involves discerning the underlying intentions and potential nuances. Understanding the *meaning* of the expression involves selecting and identifying what the verbs refer to and the object of the action. Understanding its *sense* implies knowing many more things to reconstruct its intentionality and drift (which may not always be controllable). AARON CICOUREL called it "*situated meaning*," LUCY SUCHMAN (2007) referred to it as "*situated action*," and WILLIAM CLANCEY (1997) explored it through "*situated cognition*." These frameworks underscore the importance of factors like space, time, sequence, and relational dynamics in shaping interaction. More recently, CICOUREL (2006), drawing parallels with primatologists like MICHAEL TOMASELLO and JOSEP CALL (1997), introduced the concept of "*representational re-description*." This concept refers to the evolutionary and layered construction of environments, intertwining cognitive and affective processes with social interactions and structures.

The act of extracting, formulating, articulating, and implementing a rule from and within an interaction is contingent upon the intentions of the interpreter. Turning habitual conduct into obligations, prohibitions, and permissions is a deliberate choice. In speech, numerous decisions are made, encompassing a wide array of behaviours: (i) the linguistic relationship between the interlocutors, (ii) the topics of the interaction, (iii) the personal and social relationship that exists between them, (iv) and the manner they express and convey these relationships. These components represent distinct facets of the interaction, each warranting separate consideration and analysis.

Pragmatic research and conversational analysis often depict the process as a negotiation between speakers, albeit often implicit, from which courses of action emerge. What I wish to underscore now is the similarity between constructing regulatory instruments and crafting artifacts that facilitate other actions, such as walking, cooking, or farming. Just as we engage in these activities through natural language, employing abstractions to which we assign names, and propositions, we similarly intertwine them with the act of speech. This highlights the need to differentiate between the *meaning* and the *sense* of expressions, prompting us to pose various questions. [8] Key considerations include:

---

[7] I use single quotes to indicate terms or statements, and double quotes for propositions or concepts.
[8] The relationship between the object being represented and its representation through language—whether natural, artificial, or figurative—is inherently intricate. This complexity arises from the multitude of relationships with the real world, encompassing agents, actions, and their boundaries, within which this

i. What does X *mean*? [in a context Z]
ii. What does X *signify*?
iii. What is the *sense* of X?
iv. What *function* does X perform?

Each of these inquiries directs attention towards the interplay between context, semantic content, and pragmatic function within a specific instance of term usage. The underlying thesis asserts that discerning the sense of an expression necessitates addressing all four questions, particularly by reconstructing the frame of reference and communication context under consideration.

## 4. Manage

Modal uses, beyond their prescriptive function, often entail nuanced relationships of power or authority—whether overt or subtle, explicit, or implicit—that cannot be taken for granted. These relationships are shaped and articulated through markers of acceptance, rejection, distance, irony, or sarcasm, which serve to modulate and gradate their expression. It is crucial to highlight these linguistic resources to disambiguate meanings and *make sense* of the situation.

Viewing utterances as responses or reactions, the pragmatic distance they exhibit signifies their gradual positioning on the scale. They may convey the speaker's irritation or annoyance, but they can also denote a response that goes beyond merely describing the attitude, instead outlining a planned course of action or series of actions 'triggered' by the situation at hand.

The scope of expression extends beyond mere oral or written communication, encompassing bodily movements (kinesis or elicited behaviour) as well. Conflicts often find visual expression, making body language immediately perceivable and interpretable. Emotions particularly favour this mode of response.

In human interactions, spanning micro and macro levels, whether individual or group, we observe an evolutionary response that, while manifesting in various forms across different societies, remains universal. Affectionate and aggressive relationships, akin to those observed in primates, are widespread, as are efforts to mend and restore relationships through acts of forgiveness and reconciliation. Let's examine the distribution and demeanour of the participants in the discussion depicted in Figure 1 of the document.

---

representation theoretically operates or should operate. The virtuality of rules within this framework is both cognitive and functional, as language transcends its confines to incorporate and reference various contexts. I have recently referred to this encounter of languages (natural or artificial) with the reality they inhabit, and which they also contribute to shaping, as the *impenetrability* of language (CASANOVAS, 2023). This theme has been explored in art since the era of the surrealists in the first half of the 20th century. Languages, along with sign systems, do not merely mirror reality as it is or as we perceive it.

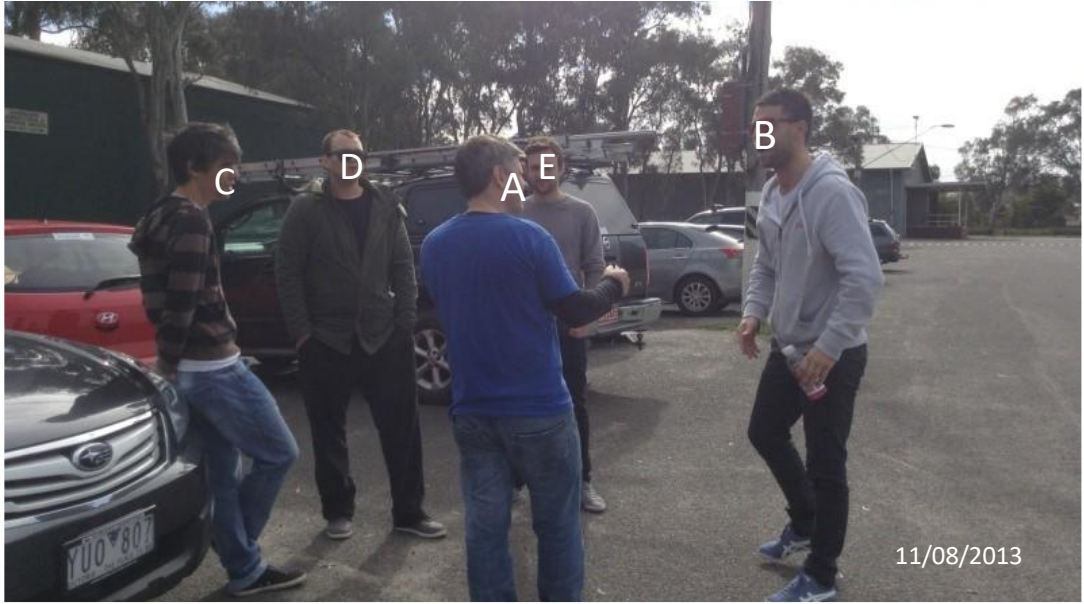

**Fig. 1**. Conflict situation (Melbourne, 11/08/2013). Original photograph by MARTA POBLET. Both she and I were present (FG). The discussion lasted about two hours. In-interaction: A-B (discussion). Passive participation: (CDE). Two groups in conflict: (AC) vs. (BDE). Participant observation (FG): out of the box.

What is significant here is not the specific data—such as who's involved, why, or the subject of the discussion—but rather the metadata that could be observed by an entirely external annotator, one who might even be ignorant of the context, focusing solely on disposition, kinesis, and body attitude to discern the participants' emotions. In the figure, albeit unconsciously, the conciliatory body disposition of A and the aggressive stance of B closely resemble, if not mirror, those described by primatologists like AURELI and DE WAAL (2000) in the conflicts of other primates. As DE WAAL (2009: p. 81) notes, facial expressions and body language convey the same emotions, and even without seeing the face, we can identify fear and anger exclusively from body position.

An analysis of linguistic interaction would be incomplete without considering the situational context, the disposition and interactive framework of the participants, and, beyond that, the composition of the groups and the defining features that institutionalize them. These elements constitute what VICTOR SANCHEZ DE ZAVALA (1973, 2007a: p.166) referred to as "quasi-pragmaticity" or "second-order pragmatics" in his psychological formulation of language theory. Emotions and their influence on the generation and filtering of information between speakers and interlocutors create a tangible behavioural envelope. While we cannot delve into a comparison of the semiotic frameworks developed by SÁNCHEZ DE ZAVALA (1973, 2007b) with the relational models of empathy crafted by DE WAAL (2000) to explain conflict resolution here, it's worth noting the meticulousness with which SÁNCHEZ DE ZAVALA sought to deconstruct the environmental components, which he termed "*transfondo*" (*background*), distinguishing them from the situational "context".[9] These are "parameters that capture the effects of the total situation in which the semiotic action takes place" (ibid. 2007b: p. 210 *et seq*.).

The cases of conflict (with micro-situations of attachment, aggression, and reconciliation) as well as the range of procedural instruments ranging from negotiation to mediation, arbitration, restorative justice, and systems adjudication courts, have been classified and studied separately by many human and social sciences, not always related to each other (CASANOVAS AND POBLET, 2007: p. 242-44, 2008).[10] It is worth noting that there are at least twenty-nine different scientific fields focusing on relational justice, all of them with their own theoretical arsenal. We could draw a table based on four major areas: (i) Behavioural research, encompassing studies on the mind, language, memory, empathy, and emotions (e.g., neuroscience and cognitive science); (ii) Social research, focusing on culture, language, excuses, and micro-situations (e.g., socio-linguistics, functional and intercultural pragmatics); (iii) Social studies, examining conflict resolution, dialogue, and reconciliation (e.g., anthropology and intercultural studies); (iv) Philosophy and law studies, delving into restorative justice and legal systems (e.g., human rights, criminology, law, and ethics). Aggression, conflict, forgiveness, peace, and reconciliation are broad social, economic, and political issues. Legal theory, criminology and judicial studies cover only a small part of the whole landscape.

5. **Regulate**

The transition to language signifies a shift in register. However, speaking is not merely about expression—*it is inherently regulatory, consistently so*. When we address someone, we implicitly propose a regulatory framework for the relationship, capable of establishing, confirming, or even sanctioning dynamics that extend beyond the literal meaning of words, sentences, or expressions. As Humpty-Dumpty famously remarked, whenever we speak, *we exert control*. Yet, it is not just linguistic meaning that we influence; we also shape the reference point and, beyond that, the entire situation, the interlocutors involved, and the potential interpreters. In essence, we propose an ecosystem. This notion will lead us to the dilemmas that arise in constructing smart legal ecosystems, which I will discuss further at the end.

However, the converse does not hold. Regulation encompasses not only speaking or formulating rules through language, but also involves *designing*. Sometimes, this design process is mediated by language, resulting in the creation of linguistic instruments such as normative systems (including legal ones), standards, or codes of conduct. Other times, regulation occurs through actions that shape the environment and guide potential behaviour of agents. Drawing an analogy with linguistic counter-performativity, we can term this *environmental counter-regulation*, wherein we construct a context, atmosphere, or environment that influences or directs individuals to behave in certain ways, without the explicit formulation of rules or mandates based on modalities of prohibition, permission, or obligation.

---

[9] SÁNCHEZ DE ZAVALA (1973, 2007b: p. 2010 and ff.) delineates these parameters, suggesting a division between those stemming from the general environment or situation, which serve as a backdrop against which semiotic interactions between interlocutors unfold, and those specifically encompassing these interactions—the semiotic situation or context. The background comprises three key aspects: (i) "Physical" conditions, (ii) The organizational framework, and (iii) The conditions of semiosis (*ibid*. p. 211).
[10] Cf. also the *White Book on Mediation of Catalonia* (CASANOVAS, MAGRE, LAUROBA, 2010-11).

The elimination of traffic lights in the design of roundabouts to avoid accidents and traffic jams, as envisioned by engineer HANS MONDERMAN (1945-2008), serves as a compelling example (see Figure 2). In this scenario, regulation is implicit within the design itself. Roundabouts do not impose rules in the same way that traffic lights do; rather, they establish a structural framework that inherently constitutes an ecosystem. This framework operates without direct prohibitions and permissions, unlike traffic lights. While certain restrictions exist—such as the requirement for drivers to navigate the roundabout by turning right or left—they are part of a broader traffic plan that encompasses both the streets and the drivers who traverse them. The concept of *shared space* underpins this approach, wherein drivers must negotiate their movement alongside others who share the road. Instead of adhering to traffic signals dictated by rules, drivers coordinate their vehicle's movements in harmony with those of others.

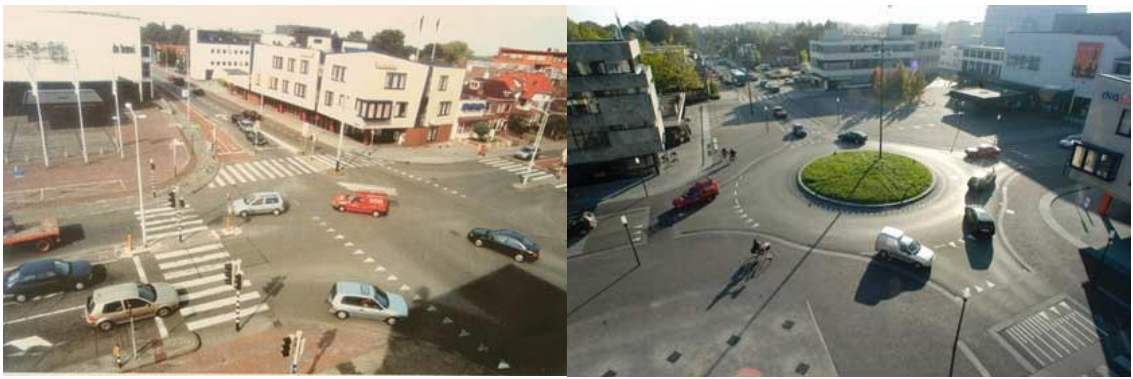

**Fig. 2**. The most problematic intersection in Drachten (The Netherlands), before and after Monder-man's solution Font: Naparsetek (2008).

The same principle applies when driving on the left in the streets of Melbourne, Australia. Here, drivers must remain highly attentive to oncoming vehicles and be prepared to stop and yield when necessary. Visual communication between drivers is continuous, often supplemented by gestures to convey understanding. In areas without roundabouts, such as intersections where right turns are required, unwritten rules come into play. In these situations, drivers may need to deviate from officially established rules, such as stopping at red lights. Instead, they may pause in the middle of the avenue *after* the light changes, waiting for a brief window to proceed before oncoming traffic resumes (see Fig. 3). This relational approach to design emerges from the interactive behaviour of agents, even if it means occasionally violating formally written and promulgated codes. These dynamics pose significant challenges for the design of *connected automated cars* (CAVs), a topic that has already begun to be studied.

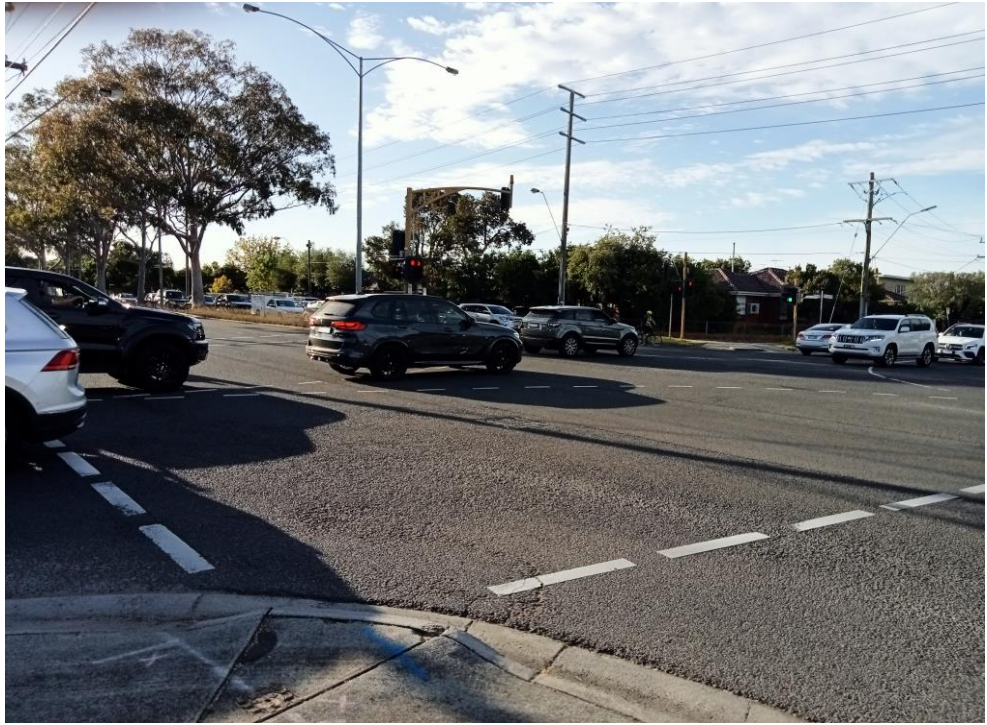

**Fig. 3.** At the intersection of Bambra and North Road in Melbourne, a distinctive driving scenario unfolds. Here, vehicles are required to wait after encountering a red light before executing a right turn, allowing oncoming traffic to pass first. This practice reflects a nuanced approach to traffic management, emphasizing coordination and mutual accommodation among drivers. Photo by P. CASANOVAS (28/02/2024).

The right turn under such conditions serves as a significant contributor to accidents in Victoria.[11] The need for sensors to learn emergent social rules has prompted the codification of drivers' gestures (GUPTA, 2016). Moreover, the right turn has been identified as a crucial use case for designing intelligent agents (COLLENETTE *et al.*, 2022). Intriguingly, the adaptation of behavior by Connected Automated Vehicles (CAVs) through rule-breaking has been proposed as a means to align with the effective rules operating on the streets, enhancing road safety and safeguarding vulnerable road users (REED *et al.* 2021).[12] These rules, whether formal (traffic codes) or informal (adaptations and changes by pedestrians and drivers), play a pivotal role (LATHAM AND NATTRASS, 2019). From a *legal liability* standpoint, however, the challenge lies in the inability to fully predict, for both manufacturers and regulators, how machine learning algorithms integrated into CAVs will behave in the future (MACKIE, 2018: p. 1318). This uncertainty underscores the complex legal and ethical considerations surrounding the deployment and operation of autonomous vehicles in real-world environments. It is one of the aspects of what is called *adaptability* in AI.

---

[11] Between 2012 and 2015, 17,6 % of accidents (Morris *et al.*, 2021: p. 8).
[12] According to Reed et al. (2021), minimizing road safety risks and ensuring transparency in the operation of Connected Automated Vehicles (CAVs) necessitates permitting these vehicles to deviate from traffic rules in instances of unexpected incidents. The authors argue against the implementation of new regulations specifically governing traffic behaviour by CAVs. Instead, they propose the adoption of *ethical goal functions* as integral components of hybrid AI systems established by governmental authorities. These ethical goal functions would guide the behaviour of CAVs, enabling them to navigate unforeseen circumstances while prioritizing safety and ethical considerations.

One way to analyse the situation depicted in Fig. 3 involves identifying the rule followed by drivers. To formalize the situation, the following steps could be outlined:

[1] If a driver (S1) wants to turn right at an intersection with a red traffic light, she (should, must, has to, necessitates):
(i) Violate the rule prohibiting passage through red lights;
(ii) Wait until vehicles approaching from the right (S2) and from the front (S3) have passed;
(iii) Execute the right turn before vehicles approaching from the left (S4) and from the front (S5) start moving when they have a green signal.

The formalization of the rule is complex. In this formulation, it is crucial to note that [1] entails the violation of another rule as a condition, which aligns with what deontic logic terms as *contrary-to-duty obligation*.[13] These instances of rule-breaking, failure to fulfill duties, or on-the-spot assumption of exceptions to the rule pose challenges that have propelled advancements in deontic logic and the logic of norms since the 1960s. Notable examples of paradoxes include situations such as: 'it is forbidden to kill, but if one kills, it must be done with as little pain as possible'.[14] However, the key point to emphasize here is that the formulation of the rule depends *on both prior textual provisions and emergent socially created situations* (as depicted in Fig. 3).

Indeed, the formulation of the rule can vary depending on the context and the perspective taken. For instance, if it is emphasized that the *only* feasible way to execute a right turn is to do so despite the red light, the functor should be framed not as *deontic* but as a *necessity ,i.e. anankastic*. Alternatively, if the rule is situated within a broader practical reasoning framework, it may be argued that no prior rule is violated. Instead, the driver proceeds through the green light and only halts *afterward* to allow other vehicles to pass before making the right turn.

The challenge of aligning Connected Automated Vehicles (CAVs) with traffic codes has been highlighted due to potential ambiguities, vagueness, and regulatory conflicts (LEENES AND LUCIVERO, 2014; PRAKKEN, 2017a, b). However, it is important to note that ambiguity extends beyond linguistic realms to encompass *pragmatic and situational factors*, as illustrated by Fig. 3. How we interpret and model the actions of drivers to derive the rules they should adhere to is a complex issue. The science of design must proactively address *contextual ambiguity*, as discussed further in Section 15.

---

[13] "A contrary-to-duty obligation is an obligation that is only in force in a sub-ideal situation. For example, the obligation to apologize for a broken promise is only in force in the sub-ideal situation where the obligation to keep promises is violated." (VAN DER TORRE AND TAN, 1999: p. 50).

[14] This is known as the 'FORRESTER's paradox of gentle murder'. Its original formulation reads: (i) Jones murders Smith, (ii) Jones ought not to murder Smith, (iii) If Jones murders Smith, then Jones ought to murder Smith gently. (i) and (iii) entail: (iv) Jones ought to murder Smith gently; (iv) entails: Jones ought to murder Smith. Cf. GOBLE (1991).

The design challenges surrounding self-driving vehicles without human drivers remain numerous and unresolved. One such challenge involves equipping these vehicles with the necessary reflexes to respond effectively in unexpected emergency situations. Humans react intuitively in such scenarios, whereas machines do not possess such innate capabilities. As RUSSELL (2023: p. 57) points out, there is currently no autonomous vehicle equipped with the understanding that people don't like to be killed.[15] Even with the application of inductive machine learning methods, the problem of transitioning from perception to decision-making persists.

The management of intelligent interactions with the environment in the creation of complex contexts constitutes the relational dimension, which encompasses both behaviour and the guidelines or patterns of regulation. The concept of "relational" holds a broader significance, reflecting the social and economic ties between agents with various roles and functions, including companies, suppliers, customers, consumers, citizens, digital neighbours, teachers, students, and drivers. It denotes the capacity to establish a shared regulatory framework characterized by reciprocity in terms of goods, services, attitudes, and actions. Consequently, "relational law" emphasizes trust and dialogue over the imposition of formal procedures or enforcement of sanctions, although these measures are not entirely excluded.

Conditions sometimes require coordinated behaviour patterns that are learned and transmitted, while in other cases, regulatory models are established to address ecological, production, or consumption challenges. ELINOR OSTROM's models, as exemplified by ADICO's *institutional syntax* and the *Institutional Analysis and Development* network (IAD), illustrate this approach (OSTROM, 2005).

From a technological perspective, we have defined *relational law and justice* as the allocation of behavioural expectations (rights and obligations) within a shared technological framework (CASANOVAS AND POBLET, 2008). Interactions among agents, as well as between agents, programs, and human-machine interfaces, create added value that facilitates bottom-up connectivity in the realms of the Web and Industries 4.0 and 5.0. Trust, security, and reliability, not just user identity, are crucial and must be integrated as characteristics of the normative models *once established*. This fosters an ecological niche where different types of technology and behaviour, whether human or artificial, converge toward common outcomes. Hence, the concept of legal ecosystems based on human-machine interactions is significant, as elaborated further later on.

---

[15] "(...) the human designer objective is clear—don't kill pedestrians—but the agent's policy (had it been activated) implements it incorrectly. Again, the objective is not represented in the agent: no autonomous vehicle today knows that people don't like to be killed." (RUSSELL, 2023: 57)

The emergence of rules and our actions concerning them presents another challenge. Philosophers like A.G. CONTE (2016: p. 87 *and ff.*), advocate for the existence of a *nomotropic space*, which serves as a functional arena for social reactions to rule content (*acting-in-function of rules*). From a different standpoint, nominalist deontic logicians, influenced by HOHFELD (1923), discuss a *space of normative positions* where logical-deductive inferences between norms are situated and operate. [16] These constitute two different epistemological approaches for modelling actions, stemming from phenomenology or analytical philosophy,

6. **Interpret**

The complexity of the relationship between language and reality is particularly pertinent in the realm of law. Jurists, particularly those specializing in public law, often presume that the law regulates a reality that must adhere to or conform with it. This initial dichotomy between those who command and those who obey, regulators and regulated, and those who possess or lack power, traces its origins to theological sources and has recurred across various cultures and political systems since ancient empires, as VOEGELIN (2001) observed.

In the philosophical and logical examination of law, the structure of rules is typically depicted as stemming from a set of antecedent conditions that yield a subsequent set of 'legal' effects or consequences, contingent upon authority or power relations. These legal effects are grounded in the negative or positive sanctions that authorities can impose to deter violations and ensure compliance. I will return on this relevant topic later (Sections 13-14).

However, the expressive function also plays a role in this domain. The pragmatic reconstruction of interaction and the interpreter's objectives are crucial for discerning the meaning and sense of expressions. In regulation, this process is both linguistic and functional, serving specific objectives. Meanings can vary and are often tied to the instrument's functionality. Consequently, the proposed or reconstructed meaning of a term or expression to convey significance in a particular situation may diverge significantly from its original meaning.

I will give an example of what JERZY WRÓBLEWSKI called the "language of jurists," to distinguish it from the "language of the law." It is a matter of framework, because it is precisely the jurists who introduce into the laws the concepts they themselves construct. But how? How do they determine their meaning?

---

[16] Cf. SERGOT (2013: p. 353 *and ff.*): "The KANGER-LINDAHL theory of normative positions is an attempt to apply the tools of modal logic to the formalisation of HOHFELD's 'fundamental legal conceptions', to the construction of a formal theory of duties and rights, and to the formal characterisation generally of complex normative relations that can hold between (pairs of) agents with regard to an action by one or other of them. The theory employs a standard deontic logic, a logic of action/agency of the 'brings it about' or 'sees to it' kind, and a method of mapping out in a systematic and exhaustive fashion the complete space of all logically possible normative relations—or 'positions'—of some given type. The article presents a generalised version of the methods and a brief discussion of its limitations as a comprehensive theory of duty and right."

In a later period, WRÓBLEWSKI and MARCELO DASCAL worked together to offer a pragmatic theory of legal interpretation because they realized that while using technical language, jurists also had to use concepts from natural languages to start from a common base and have shared knowledge. They questioned what we can call the expansive interpretative intent of jurists, expressed through the classical principles of *clear non sunt interpretanda* and *interpretatio cessat in claris*. And they also differentiated between "assuming" the meaning of a legal text and "interpreting" it. Not all texts are interpreted. Clarity is treated "as a pragmatic feature of the use of legal text in a given situation rather than as an absolute property of the text itself".[17]

It is quite interesting that WRÓBLEWSKI and DASCAL grounded their pragmatic theory in relational semiotics rather than (i) the completion or completeness of meaning by contextual variables; (ii) or the use of indirect strategies, as is the case with PAUL GRICE's maxims to explain the difference between the *speaker's meaning* and the *ostensive utterance-meaning*. They advocated for *heuristic abduction*, initially fallible but functional. Later, they developed a theory on rational regulation (or *rational ruler*) along the lines of the pragmatic theory of argumentation, separating in legal discourse the realms of legislation (*law-making*), application (*law-applying*), and doctrine or theorization (*law-describing*).[18] Essentially, they aimed for a general theory of interpretation that respected the characteristics they attributed to the pragmatic rationality of legal language.[19]

However, this is an analytical description that corresponds more to the norms [*Normen*] as conceived by German jurists from the second half of the 19th century. Law, until it began to be conceived as an internal and external system to the Enlightenment—*sistema juris, sistema legis*— consisted of a set of practices, behaviours, concepts, discourses, and topics that drew from classical logic, topics, and rhetoric developed from the reception of Justinian Roman law during the Middle Ages, the Renaissance, and the Baroque period. Naturally, it was linked to the emergence of state forms as well as transactions in the market and social and family relations.

Law as knowledge is related to practical philosophy and the application of problem-solving knowledge (such as medicine and engineering), but as language and a universe of language, it has always seemed to me like the reverse side of the mirror that contains everything in front of it and reflects both its image and its reverse from any angle. Therefore, like in MAGRITTE's[20] painting, its image is unfaithful because it does not merely reproduce what is there but continually recreates it, sometimes surprisingly, because the so-called "legal reality" absorbs any other reality and shapes it and gives it content based on the cultural, political, and institutional framework it adopts. In my view, from this perspective, it is a dimension of the social that regulates and assigns a specific value to any imaginable entity, action, or behaviour.

---

[17] They differentiated between: (i) *Sensu largissimo* (cultural) interpretation; (ii) *Sensu largo* interpretation (linguistic, according to the accepted rules and uses of a natural language); (iii) *Sensu stricto* interpretation (problematic: the case of metaphors and ambiguity). Cf. DASCAL AND WRÓBLEWSKI (1989).
[18] Ibid. p. 434. DASCAL AND WRÓBLEWSKI (1991) included justification into their pragmatic interpretative formulation: "N the legal rule R in the legal language $L_1$ and/or in context Ci has meaning M according to the directives of interpretation $DI_{11}, DI_{12}, \ldots, DI_{n1}; DI_{12}, DI_{22} \ldots Dl_{n2}$, and the evaluations $V_n, V_{,i} \ldots V_{ni}$ necessary for the choice and use of the directives".
[19] "In spite of its fuzziness and context-dependence, legal language ensures sufficient uniformity to allow for regular and reliable "translations" or interpretations of the rational law-maker's utterances into the interpreter's utterances." (*ibid*.)
[20] I am referring to Magritte's painting *La reproduction interdite* (1937), preserved in the Museum Boijmans-Van Beuningen in Rotterdam. I had the opportunity to analyse it in detail in CASANOVAS (2023).

For this reason, philosophers have sought some kind of ontology or structure that can account for entities, actions, relations, and social forms across any human society. They operate as *interpretive schemes* that allow defining the scope of what is considered 'legal', distinguishing legal, moral, and social norms, and establishing the criteria of interpretation applicable to attribute meaning to concepts.[21] In a more metaphorical way, but just as effective, a writer like CAMILLERI advocates *wool rules* over *iron ones*: The rules would be understood as a wool sweater that can be adapted, i.e. *accommodate*, depending on who wears it.[22] We will return to this in the following sections, because as argued by DASCAL, WRÓBLEWSKI, and SANCHEZ DE ZAVALA, the quintessential form of pragmatic inference that describes the texture of this woollen sweater is *abductive*.

## 7. Signify

However, perhaps the elaboration of a general theory and equally general principles may not be the path forward now. To *signify* is to reconstruct not just meaning but sense. Assumptions about the general structure of law and legal instruments have been affected by the digital society and the construction of new regulatory instruments. I believe it would be beneficial to lower the level of abstraction and make weaker presuppositions. The first step is to attend to the complexity of the context and pay attention to the 'subjective' or 'intentional' sense of statements. This does not imply an abandonment of theory. On the contrary, it is a further step towards the development of theories in the normal sense, i.e., ones that are susceptible to formal and empirical testing, i.e., containing explanatory hypotheses that can be subject to validation.

There is a very nice example that the Roman law scholar JUAN MIQUEL analysed in *Aenigma*, the Inaugural Lesson of the 1975-76 Course at the University of La Laguna, to show the difference between the reconstruction of the meaning of a sentence and the sense it can have in its context.

---

[21] Perhaps the most well-known are: (i) JOHN R. SEARLE 's schema of the construction of social reality, (ii) A.W. HOHFELD's two matrices of 'fundamental jural concepts', (iii) ALF ROSS 's inferential conception of law, (iv) STEPHEN TOULMIN 's argumentative model, (v) the distinction between primary and secondary norms by HANS KELSEN, HERBERT HART,and ALF ROSS, (vi) the so-called "procedural conception of law," based on the concept of practical rationality defined by ROBERT ALEXY, ALEKSANDER PECZENIK, AULIS AARNIO, and later, JÜRGEN HABERMAS. The latter two schemes are also considered "theories of law" because they delineate specific properties of legal systems such as *validity* (legality), *efficacy*, *effectiveness*, *coercibility*, and *justice*.

[22] Cf. CAMILLERI (2004: pàgs. 234-235). "Io so benissimo che ci sono le regole! [...] Sono come il maglione di lana che mi fece zia Cuncittina. [...] Quanno avivo una quinnicina d'anni, me' zia Cuncittina mi fece un magliuni di lana. Ma siccome non sapiva usari i ferri, il magliuni aviva ora maglie larghe che parivano pirtuse ora maglie troppo stritte, a aviva un vrazo più corto e uno più longo. E io, per farmelo stare giusto, doviva da una parte tirarlo e d'altra allíntarlo, ora stringerlo e ora allargarlo. E lo sai perchè potiva farlo? Perchè il magliuni si prestava, era di lana, ..non era di ferro".

MIQUEL focused on a seemingly innocuous phrase from a Roman jurist of the 2nd century AD, CELS [PUBLIUS JUVENTIUS CELSIUS], commenting on and opposing the theses of a slightly earlier jurist, TUBERO [QUINTUS AELIUS TUBERO], regarding the elements that composed the movable property that could be inherited. The phrase, later included in the *Digest*, is very simple: *Tubero hoc modo demostrare supellectilem temptat*, i.e., "Tubero attempts to define the movable property [or household effects] in this manner".

Apparently, CELSIUS treats TUBERO with the deference that is customary among jurists— "although I greatly appreciate TUBERO's authority"....—. But MIQUEL demonstrates that CELSIUS's intention is exactly the opposite of what he says, diminishing its pomposity, because TUBERO came from the field of rhetoric, not law, and according to CELSIUS, he lacked the necessary precision that is the mark of a jurist. Thus, if we understand (i) *demostrare*, (ii) *supellectilem*, and (iii*) modus* as rhetorical technical terms derived from Greek, i.e., (i) *ostentation* or exhibition, (ii) *ornament* (to move), (iii) *rhythm* or style, their use renders the meaning of the phrase enigmatic.

The sense of *TUBERO hoc modo demostrare supellectilem temptat* is not then apparent but hidden in the choice and use of terms: "TUBERO attempts to display his rhetorical artifice with this rhythm (or style)." And if one considers that *tuber* also means 'affected,' 'hypocrite', 'pompous', and 'intrusive,' we already have the complete reproach: "Tubero is a pretentious newcomer who replaces analysis with rhetoric." A snob, an intruder, a pedant, in short. This already implies a prescriptive modality: "Rhetoric [to impress an audience] must be excluded from legal knowledge". The jurists of the time understood this immediately. However, to do this from our perspective and not think that the meaning of the phrase is unique, we need to reconstruct the context and dig a little deeper.

To prove it, i.e., to validate this interpretation, MIQUEL also proceeds to a meticulous *colometric* analysis of the rhythm and style of CELSIUS 's text, which parodies and thus mocks the rhythm with which TUBERO arranged the enunciation of the text [2]:

[2]  *instrumentum quodam // pater familiae*
*rer(um) ad cottidianum usum // paratarum*
*quod in aliam speciem // non caderet*
*ut verbi gratia // pen(um) argentem vestem*
*ornamenta instrumenta // agr(i) aut domos*[23]

MIQUEL was inspired by WILLIAM EMPSON's idea in *Seven Types of Ambiguity* (1930) that every concept in language includes multiple latent meanings with which one can play, whether in poetry or in law. EMPSON referred to this method of breaking down units of meaning as *close reading*; in computational linguistics today, it is called *disambiguation* (or *word-sense disambiguation*).[24]

---

[23] A specific tool // items for daily use of the head of the household // belongings that would fit in no other category like, for example, // a fine silver dress ornaments, tools // fields or houses.
[24] Vid. on the relationship between the two methods, GAVIN (2018).

The Roman law scholar did not stop here, but also applied the idea of *validation*. In history and social sciences, interpretations must be validated to be considered true; and for this reason, he conducted further analyses, measuring the Latin rhythm of statements, applying metrics to the completion of sentences (colometry), and seeking quotes from contemporary authors describing the way of thinking and writing of CELSIUS. TUBERO wrote during the time of AUGUST (63 BC-14 AD), extolling the empire and feeling self-importance; CELSIUS, during the time of ADRIAN (76 AD-132 AD), separating political and legal objects. ADRIAN was known precisely for the great administrative reforms of the state and the professionalization of jurists. These, therefore, are the different interpretations of *Tubero hoc modo demostrare supellectilem temptat*, enunciated by CELSIUS and reconstructed by MIQUEL (my paraphrase):

[3] TUBERO attempts to define the movable property [or household effects] in this manner.
[4] TUBERO attempts to display his rhetorical artifice with this rhythm [or style].
[5] As his name indicates, TUBERO is a pedant who is replacing legal analysis with rhetoric.
[6] Rhetoric [to impress an audience] must be excluded from legal knowledge.

[3] and [4] are two translations of the specific meaning of the Latin statement according to the possibilities of signification of the concepts it contains, referring to the jurist and what he tries to do. The double meaning of the sentence is an effect intentionally created by CELSIUS. [5] is an intentionally added sense that performs an illocutionary act of mockery, playing with the different linguistic uses of TUBERO, to reinforce the interpretation of [4]. [6] is an implicit prescription that follows from CELSIUS's theoretical position on the functions of law and jurists. It reveals at the same time as it explicates and justifies CELSIUS's intentions.

Jurists' use of language should be perspicuous and precise. It is like the moral of the story. *Rhetoric [to impress an audience] must be excluded from legal knowledge.* Therefore, we see that a complete clarification of these six words requires the reconstruction of the sense of the statement through a layered analysis that takes into account all relevant elements for understanding not only the utterance (including morphology and rhythm), the statement, and its intentional effects, but also the speakers, interlocutors, and recipients (audience) that are part of the context. Since the context is evolutionary, moreover, we can add the unintentional effects: CELSIUS's own historical function as a jurist at the time of the reconstruction of the imperial administration. I was not the only one impressed by what JUAN MIQUEL had achieved with this work.[25]

---

[20] MIQUEL was deeply interested in the pragmatics of languages. I remember several discussions with him in the late 1980s and early 1990s about pragmatic markers and the value of certain particles in Eastern languages—e.g., "ta" in Japanese and Korean at the end of a sentence as emphasis or a sign of politeness. This was when he was learning Japanese to give some lectures on Roman law in Japan ("I will surprise them"). He argued that the meaning of utterances (not just statements) required incorporating their pragmatic value. The linguistic approach, especially the colometric analysis of rhythm related to content and the reconstruction of meaning in *Aenigma*, impressed ROMAN JAKOBSON, who sent him an enthusiastic

# II Second part: Computing

In this second part, I will resume the discourse on ambiguity that I began in the first part. The first part has addressed the epistemic (pragmatic and cognitive) foundations of meaning (as signification) in law. The second part will undertake computational analysis and the construction of regulatory systems, after examining some positions in legal theory. I will only recall the sequence of linguistic activities with which I have organized this second part: 8. *Elucidate*. 9. *Disambiguate*. 10. *Generate*. 11. *Regenerate*. 12. *Comply*, 13. *Execute*. 14. *Contextualize*. With a final section on *Reexpress* (15) where I will make some conclusions explicit.

## 8. Elucidate

In the philosophy of ordinary language, the operation of 'discovering' or 'clarifying' the meaning of concepts through analysis is called "*elucidation*." Especially in the analytical jurisprudence by H.A. HART (1956-57, 1961, 1994, 2012), elucidation has been extensively used as a "method" – the "analytical", "philosophical" elucidation of reflection on the operative conditions of concept usage: (i) selecting some examples, (ii) figuring out different scenarios of their use, (iii) identifying standard uses, (iv) specifying how they differ from each other, (v) defining what they have in common, (vi) comparing them with the uses of terms and expressions in natural language (English, in his case) outside the specific legal context where the concept to be "elucidated" operates. This approach helps reduce the ambiguity of classical legal concepts such as 'rights', 'obligation', 'duties', or even the concept of law itself, i.e., what does 'Law' mean. [26]

This technique has both pros and cons. Among the advantages, it has the capacity to discriminate between different concepts in relation to their operative functions. For example, between "social rules and mere converging habits of conduct" (HART 1992: 12), or between different types of implementations of the same rules.[27] However, there are also disadvantages. Mainly: (i) the

---

congratulatory letter, which he showed me, and I was able to read. It was a large white card, unfortunately lost today.

[26] "The technique I suggested was to forego the useless project of asking what the *words* taken alone stood for or meant and substitute for this a characterization of the function that such words performed when used in the operation of a legal system. This could be found at any rate in part by taking the characteristic sentences in which such words appear in a legal system, *e.g.,* in the case of the expression "a right," such a characteristic sentence as "*X* has a right to be paid *Y* dollars." Then the elucidation of the concept was to be sought by investigating what were the standard conditions in which such a statement was true and in what sort of contexts and for what purpose such statements were characteristically made. This would get away from the cramping suggestion that the meaning of a legal word is to be found in some fact-situation with which it is correlated in some way as simple and straightforward as the way in which the word "chair" is correlated with a fact-situation and substitute for this an inquiry into the job done by such a word when the word was used in a legal system to do its standard task." (HART, 1956-57: p. 961).

[27] "[…] jurists seemed often to have ignored the fact that rules are not only susceptible of being *obeyed* or *disobeyed* but of radically different operations. For example, when rights are claimed, a rule is *invoked;* when a legal power is exercised a rule is *acted on;* when a particular case is found to fall within the scope

disconnection from any empirical method of validating the results of the analysis; (ii) and the need to start from an already established framework of typical uses that delimit the characteristics of a specific field or domain; (iii) the assumption that legal theory alone is sufficient to constitute itself as a method through philosophy of language, without recourse to what KUHN called "normal science"; (iv) the assumption of the state as the realm of the legal.

In essence, it is assumed that uses already exist and are known by legal operators—the legal profession—, and that philosophical analysis of language (not linguistic, philological, or sociological) reveals and brings them to light. "Elucidation" as a method thus leads to an *assumed elicitation* of meaning from those who habitually use the concepts —lawyers, judges, prosecutors.[28] For practicing it, it is not apparently required any other method of knowledge acquisition, i.e., it is carried out without any guarantee outside of their own experience. In my opinion, this covers up the consequences of what we might call the 'common sense' of the jurists, which inevitably emerges through their formulations. Furthermore, it also inherits biases: *The "elucidated" language of law is the 'official' language of law*.

If one compares this approach by HART with that of JUAN MIQUEL, for example, the first surprising thing is the ease with which the philosopher accepts that mere conceptual (philosophical) analysis can constitute an empirical method, even while placing itself on a descriptive plane. MIQUEL moves beyond the language of jurists to explain and reconstruct its meaning, and then verifies it. HART, on the other hand, is content to refine its meaning without the need for any subsequent verification.

It is quite interesting to trace the posthumous analysis of HART's *Postscript* (1992), carried out (and also posthumously published, like the *Postscript*) by his successor in the Oxford chair and main interlocutor and critic, RONALD DWORKIN (1931-2013). DWORKIN reiterates the arguments put forward in *Law's Empire* (1986): (i) jurisprudence (in the sense of legal theory) is also normative because it explicates the most practical concepts that occur at another level and participates in the internal game of interpretation; (ii) therefore, it cannot be "semantic" or "archimedean" (in the sense of mere description or theorization of law from the outside); (iii) the "rule of recognition" elaborated by HART as the ultimate criterion of 'legality' or 'validity' cannot hold a universal scope; (iv) law (including propositions about or stemming from law) occurs "within" law; (v) the truth conditions of basic legal propositions (i.e., that "p is law") are always relative to a specific legal system and legal operators who interpret, in a strong sense, what they consider

---

of a general rule, a rule is *applied.* These and many other different relationships to rules need elucidation." (HART, 1956-57: p. 958)

[28] The *Cambridge Dictionary* defines 'elicitation' as "the process of getting or producing something, especially information or a reaction." I use this term similarly to how it is used in anthropology, sociology, sociolinguistics, and cognitive sciences to gather information through interaction with speakers or informants. HART (and many analytical legal philosophers after him) dispense with these steps because they assume that they already grasp the basic meanings and conceptual uses of the lexicon of a particular community of speakers.

true or false. DWORKIN takes these arguments to their conclusion—law can only be "known" interpretively, and legal "facts" have a different nature than purely natural ones.

> […] the fact that law's character is revealed *by* interpretation rather than scientific, or natural, investigation is a signal that law is not a natural kind. We can generalize that claim to cover social institutions generally. Plainly there is no physical or social or psychological fact that fixes the nature of marriage, for example. Suppose it is true that there has never been a society in which homosexual unions of anykind have been recognized as marriages. It would not follow that it is of the essence of marriage, as a social institution, that it be heterosexual. Any claim that this is part of the essence of marriage - that a union of homosexuals, whatever else it might be, cannot be a marriage - is an interpretive claim, and must be assessed in that way. (DWORKIN, 2017: p. 2102-03)

The "legal paradigms"— the sense that legal operators attribute to what is taken from the law to resolve cases ("of law is taken to be settled"— are broadly accepted and shared by legal operators and change over time (*ibid*. p. 2123). They use a *criterial semantics*, i.e., more or less equivalent criteria, to reach agreements on the truth or falsity of legal propositions (*ibid.* p. 2112). In any case, the values, principles, and any moral standard that one may use are interpretive and internal to the practices and procedures of legal professions.

It is noteworthy that, whether in HART's positivist (external) version or in DWORKIN's more hermeneutic (internal) version, there is an occlusion where the scope is defined by their operators, their languages, and their practices, outside the wider society that accommodates them and which, at times, struggles to follow, understand, and accept their techniques. Power, authority, violence, ultimately, continue to constitute a set of exogenous variables that must be taken into account in the analysis of law.

If there is an inside/outside or vice versa, it is difficult to determine because it seems that rather than following the model of ARCHIMEDES, jurists follow that of KING MIDAS: they turn everything they touch into 'legal'. Any word, concept, or statement in natural language, then, ends up being ambiguous until there is a 'legal' concept that redefines it. We could call this phenomenon the *double loop* of legal language. To the natural ambiguity of the language it uses is added the ambiguity of the terminology it proposes. A question that might initially seem strange in the world of normal science, such as 'What does it mean the concept of Artificial Intelligence from a legal point of view?' is a perfectly acceptable question that no longer surprises experts.[29]

---

[29] This problem has been raised again due to the recent enactment in the European Community of the so-called *Artificial Intelligence Act*, approved by the Parliament on March 13, 2024. Finally, an extensive understanding of the term has been adopted, without distinguishing phases, without distinguishing phases, including practically any information system containing degrees of autonomy: "Art.3.1- AI means a machine-based system designed to operate with varying levels of autonomy, that may exhibit adaptiveness after deployment and that, for explicit or implicit objectives, infers, from the input it receives, how to generate outputs such as predictions, content, recommendations, or decisions that can influence physical or virtual environments." The definition (several times updated) by Stuart Russell, Karine Perset, and Marco Grobelnik for the OECD distinguishes between the phases of construction and use: "An AI system is a machine-based system that can, for *explicit or implicit* objectives, infer, from the input it receives, *how to generate outputs such as* predictions, *content*, recommendations, or decisions that *can* influence *physical*, or virtual environments. *Different* AI systems vary in *their levels* of autonomy and *adaptiveness after Deployment*." Cf. https://oecd.ai/en/wonk/ai-system-definition-update .

It is not enough to ask: 'What does 'artificial intelligence' (AI) mean?' One asks for its *legal meaning*, thus opening the door to its "elucidation" as soon as the term is introduced into what is deemed a 'legal text'. Echoing WRÓBLEWSKI, the blending of the language of law and language of jurists still is a difficult subject of analysis.

### 9. Disambiguate

We can argue that the issue of linguistic ambiguity is inherent to legal technique and law, closely linked to some of the schemes I have mentioned in Section 5 (note 25). It is worth considering it as it is. I am referring to the possibility of conceiving it not from the point of view of reference, which gives rise to the multiple techniques of disambiguation and interpretation, but from the point of view of cognition and language, which allow us to understand it as the set of operations that, like a hinge, modulates the way we represent the world. I refer here not only to meaning but also to sense, and therefore propose this holistic perspective that considers the situation, the "background" (*transfondo*) referred to by SÁNCHEZ DE ZAVALA, i.e., the interlocutors, agents, and organizations, in the sequence of using legal instruments in specific situations, contexts, environments, and scenarios.

Cybernetics and cognitive anthropology have pointed it out stemming from the notion of *cognitive decoupling*, i.e., the detachment or dissociation that occurs at the origin of abductive reasoning and what, lacking a better term, we call creativity or *plasticity of the mind*. It can be defined as "the ability to distinguish between assumption and belief, and to perform mental simulations to make choices" (LESCHZINER, 2019: p. 184). I will give some examples.

Since the 1950s, there has been a proliferation of attempts to apply logic to law. Since then, there has been a veritable explosion of modal logics for representing normative systems. The so-called non-monotonic logics allow us to conceive the notion of consequence based on the possibility of modifying premises with the addition of new information.[30] These are called *defeasible logics*, which have given rise to contemporary theories of argumentation and a series of *non-standard deontic logics* (GABBAY *et al*. 2013, 2021). Their use allows the configuration of *schemes* that can be applied for calculating legal effects, stemming from the formalization of starting conditions. This has in turn led to the application of computational languages and, as we will see, the use of Artificial Intelligence for the understanding and ultimately the creation of recursive legal systems. These systems are quite useful for normative systematization as they are capable of offering solutions to legislative and jurisprudential saturation, which seems to be the main characteristic of the hyper-regulated public space in which we find ourselves today.

---

[30] "The term "non-monotonic logic" (in short, NML) covers a family of formal frameworks devised to capture and represent *defeasible inference*. Reasoners draw conclusions defeasibly when they reserve the right to retract them in the light of further information. Examples are numerous, reaching from inductive generalizations to reasoning to the best explanation to inferences on the basis of expert opinion, etc. We find defeasible inferences in everyday reasoning, in expert reasoning (e.g. medical diagnosis), and in scientific reasoning." (STRASSER AND ANTONELLI, 2019)

In many of these logics, 'rights' and 'duties' are defined based on functors that refer to the permissibility, obligatoriness, or prohibition of actions that delimit the components of the initial conditions (or premises). Thus, 'X is permitted' is considered equivalent to 'having the right to do, perform, or bring about X'. Systems of rights and systems of norms are considered equivalent.

Regarding ambiguity specifically, there have been recent attempts to formalize the "logic of ambiguity" with three specific hypotheses: (i) linguistic ambiguity can be precisely defined and is not a vague notion (there are no boundary cases); (ii) in explicit reasoning with semantic ambiguity, we must always take into account the parameter of *trust*[31]; (iii) ambiguous propositions do not have the same properties as explicit or unambiguous ones, i.e., they are "second-class", and do not have their own semantics in terms of truth/falsity, but only jointly with other propositions that do possess it (WURM, 2023).[32]

However, despite the relevance of the mentioned logics, ambiguity of natural language remains an irreducible feature of the operations that can be carried out. It is, once again, about *sense-making* through the use of natural language as different from the *production of meaning* in language. Making sense of a normative situation in a specific context does not necessarily imply the materialization of the normative conditions that can be delimited in the abstract from the elucidation of the meaning of the concepts applied. *Executing a rule is not the same as interpreting its meaning in its application* (See below, Section 13). This is one of the problems of computational languages when they have to be effectively applied or "activated" in specific social contexts. It is a delicate issue because there have been numerous scandals precisely due to the poor application of information systems with disastrous consequences.[33]

The assignment of meaning in a given context is a very open game, which depends on the knowledge of the interpreter and what she selects to give meaning to the elements, the order and set of what she interprets. Terms, arrangement, visual space, distance, when not directly the written text, matter, because they implicitly trigger the learned rules that make (rather than give) sense to the text itself or to the images. It is a process of "decontextualization" and "recontextualization" that has an institutional character, *institutional framing*, to put it in terms of CICOUREL (1992) and JACOB MAY (2003).

This mix of disorderly accumulation of information also occurs in decision-making and in the shaping of public policies. We cannot assume that information or knowledge representation is always communicated in an understandable manner. Generally, it is not the case.

---

[31] "In particular there is not and there cannot be the logic of ambiguity: we always have to set the parameter +/-trust" (WURM, 2023).
[32] Cf. also Note 3 on this point.
[33] I had the opportunity to present some of these disregarded applications by administrations and their negative effects in CASANOVAS (2020).

From a political science perspective, one of the dominant approaches takes an eclectic view that attempts to capture the multiplicity of elements that must be considered to coordinate public policies in a situation of uncertainty, where government departments and agencies do not have all the information, and where decision participants have a partial representation of others' representations. This approach is the *Multiple Streams Framework* (MSF), developed by JOHN KINGDON in 1984 and reformulated in 2011, specifically based on ambiguity.[34]

The MSF has three presuppositions: (i) policymakers face significant time constraints, i.e., they cannot address all problems, they must use heuristics to design and implement policies, and they do not seek the optimal but rather what is *satisficing or good enough*; (ii) means and ends, problems and solutions, are generated independently of each other; (iii) *ambiguity permeates the entire process*.[35] This perspective thus assumes an "organised anarchy" that does not correspond to strictly defined models of rationality. "*Thus, policy problems, policy solutions, and political conditions shift constantly and without clear linkages to the others*" (HOEFER, 2022: p. 2).

Public policies related to COVID, immigration, or homelessness often result unexpectedly from multiple decisions and groups with changing intentions and objectives in a process that evolves over time. In each choice situation, policymakers respond to the main questions of attention, search, and selection under conditions of ambiguity:

> *Ambiguity* refers to having many ways of thinking about the same circumstances or phenomena. In contrast with *uncertainty*, which may be reduced by collecting more information, more information does not reduce ambiguity. For instance, more information can tell us how Covid-19 is spread, but it will not tell us whether Covid is a health, economic, educational, or civil liberties issue. Therefore, *we often do not know what the problem is*. […] (ZAHARIADIS, 2023: p. 2-3)[36]

---

[34] "The multiple streams framework (MSF) is a lens that was developed to explain United States (US) policy under conditions of ambiguity. It draws insight from interactions between agency and institutions to explain how the policy process works in organized where there is a shifting roster of participants, opaque technologies and individuals with unclear preferences." ACKRILL *et al*. (2013: p. 871).

[35] "The lens makes three assumptions. First, policy-makers operate under significant and varying time constraints. In practice this means (a) they cannot attend to all problems, (b) they must use heuristics to get things done, and (c) they must accept outcomes that satisfice rather than optimize. Second, means and ends, solutions and problems are generated independently of each other. The implication is that information is vague, consequences are uncertain, and 'there appears to be no satisfactory way of determining an appropriate set of means or ends that would obtain sufficient agreement among a diverse set of stakeholders'. Political conflict is endemic and issues are frequently settled by activating certain frames as EU actors move in and out of the process. Third, ambiguity permeates the process. Most actor preferences are opaque and not well defined; organizational technology is only partially comprehensible; participation is fluid. Information and institutions are not value-neutral. As a result, the process is open to political manipulation biased in favour of those who generate information, control access to policy venues, and synchronize or exploit group, national and institutional timetables." (ACKRILL *et al*. op. cit: p. 872)

[36] It follows: "Viewing the policymaking process as essentially a series of linked choices, MSF argues that choice depends on the coupling of three streams – problems, policies, and politics – by policy entrepreneurs during open policy windows. Each stream is viewed as an independent variable that takes several values over time. For example, the unemployment rate as a problem varies over time. Because each stream obeys its own dynamics, the MSF talks about streams being ripe for coupling, meaning that elements – problems or solutions – have become viable candidates for policy consideration. The argument is interactionist in the sense that choice does not result from adding solutions to problems but how they interact with one another over time." (Zahariadis, 2023: p. 2-3)

## 10. Generate

I will approach this section from the perspective of *Natural Language Processing* (NLP), generative AI and the so-called *Large Language Models* (LLM). In NLP, a few years ago, engineers began to classify types of ambiguities into categories, so that they were more tractable.[37] They did it in a practical (not linguistic) way, to identify bugs that could cause problems. MASSEY *et al.* (2014) singled out six types: (i) lexical, (ii) syntactic, (iii) semantic; (iv) vagueness; (v) incompleteness; and (vi) referential. They could be reduced to four, if the vagueness is removed and levels are differentiated in the context: lexical, semantic, pragmatic, and contextual.[38]

As we could imagine, in generative AI, training models with cases that present a greater degree of specificity is better than using a higher degree of abstraction.[39] *Prompts*, i.e. questions that must be asked to train the system, must not only be accurate and precise, but also reduce ambiguity as much as possible. In introducing CLAM, the researchers noted that LLMs rarely ask for clarification when faced with ambiguous questions.[40] This is an area of research that is already starting to be hammered out, but my impression is that there is still a lot of manual work in the initial process.

In the case of law, articles on *law smells* have recently begun to be published. In engineering, a *code smell* is a problem in the code that indicates there is a bigger problem underneath.[41] Examples: duplicate sentences, 'long' elements, too large reference trees... And among them the authors place *ambiguous syntax*.⁴² For the reasons I mentioned, I'm not sure I entirely agree. In any case, the reason for including ambiguity as a kind of bug or glitch cannot be that it "leaves room for interpretation".

---

[37] "*Lexical* ambiguity refers to a word or phrase with multiple valid meanings. *Syntactic* ambiguity refers to a sequence of words with multiple valid grammatical interpretations regardless of context. *Semantic* ambiguity refers to a sentence with more than one interpretation in its provided context. *Vagueness* refers to a statement that admits borderline cases or relative interpretation. *Incompleteness* is a grammatically correct sentence that provides too little detail to convey a specific or needed meaning. *Referential* ambiguity obre karefers to a grammatically correct sentence with a reference that confuses the reader based on the context provided." (MASSEY et al. 2014: p. 84) .

[38] "Ambiguities in NL requirements are normally classified into four main categories : *lexical*, i.e., the terms used have unrelated vocabulary meanings; *syntactic*, i.e., the sentence has more than one syntax tree, each one with a different meaning; *semantic*, i.e., a sentence can be translated into more than one logic expression; and *pragmatic*, i.e., the meaning of the sentence depends on the *context* in which it is used. The term *context* is used as a general concept, which includes different levels: (1) those entences immediately preceding and following the current one, (2) the other sentences placed in other sections of the document, (3) the *domain knowledge* of the subject reading the requirement (the reader), (4) the reader's *common sense knowledge*, and (5) the reader's *viewpoint*." (FERRARI *et al.*, 2018: p. 32),)

[39] "With regard to specification abstractions, higher-level requirements or specifications are often distinct from lower level specifications through the allocation of further structure and behavior within a defined boundary to satisfy one or more higher-level requirements. That is, the lower-level the specification, the more well-defined the architectural and programming constructs become. Indeed, there would be more ambiguity and difficulty in defining higher-level specifications for code synthesis, as the algorithm would need to implicitly derive an internal set of "lower-level" specifications before synthesizing the corresponding code solution." (CHEN *et al.*, 2021)

[40] CLAM és "a framework that can significantly improve language models' question answering performance in a setting we describe as *selective clarification question answering*. The framework involves: identifying ambiguous questions, prompting the model to resolve ambiguity, and answering the disambiguated question." KUHN et al. (2023).

[41] "Def. 1. A law smell is a surface indication that usually corresponds to a deeper problem in a legal system." (COUPETTE *et al.*, 2023: p. 339).

Detecting errors does not present problems, but to modify or fix the legal text because it must be interpreted is to repeat the same problem of some formulations of *Rules as Code* —not all— which aim to make the law immovable through "official" interpretations and try to reduce or eliminate the role of lawyers and legal interpretation through programming.⁴³ We already had occasion to be critical of this perspective, when the OECD popularized it in a Document entitled *Cracking the Code: Rulemaking for Humans and Machines*⁴⁴ (CASANOVAS, HASHMI *et al*., 2020).⁴⁵

A somewhat complementary but opposing proposal is that of JOHN NAY and his team at Stanford called *Law informs Code*. In this case, they assume that the legal technique operates through AI methods, applying prompting and fine-tuning techniques to carry out the training of LLM models applied to law. The premise is that generative AI recursive language models can assist practitioners and citizens in a wide variety of cases, from contract analysis to the prediction of solutions. Doing this, the access to justice is democratised, because it reduces the costs and complexity of access to the law. ⁴⁶ It must be said that this is not about automatically applying the law, but about modulating it in an iterative process that starts with the training of the models and can be extended over time:

---

⁴² "*Ambiguous syntax* is a legal adaptation of the software engineering code smell *mysterious name*, with pinches of *repeated switches* and *speculative generality*. Lawyers smell it, inter alia, when they litigate over the meaning of commas or argue about whether an *or* is inclusive or exclusive. More formally, ambiguous syntax is the use of logical operators (e.g., *and*, *or*, and *no(t)*), control flow operators (e.g., *if*, *else*, and *while*), or punctuation (e.g., commas and semicolons) in a way that leaves room for interpretation. Ambiguous syntax is problematic because it creates legal uncertainty, which is often removed through costly lawsuits and sometimes leads lawmakers to adopt *mathematically redundant syntax* like *and/or.*" (ibid.)

⁴³ "A digital twin of a citizen is a digital representation of an individual. […]. Governments are developing digital twins of citizens to monitor the environment citizens live in and address health, safety, travel and social media impacts on society. The spectrum of complexity of the models and tools can help governments make better decisions for monitoring and supporting patients, prisoners, passengers, or the elderly. Some governments, such as China's, are building a scoring methodology. Aggregated citizen twins can help map broad patterns and drive resource allocation. […] By implementing MRL [Machine-readable Legislation], the room for interpretation of legislative or executive intent is eliminated from the process, instead making the law that is passed the same as that which is implemented [my emphasis]." (GARTNER 2021).

⁴⁴ https://www.oecd.org/innovation/cracking-the-code-3afe6ba5-en.htm

⁴⁵ Both documents are available at: https://www.oecd.org/innovation/cracking-the-code-3afe6ba5-en.htm and https://zenodo.org/records/4166115 .

⁴⁶ "Law-making and legal interpretation convert opaque human goals and values into legible directives. *Law Informs Code* is the research agenda embedding legal processes and concepts in AI. Like how parties to a legal contract cannot foresee every potential "if-then" contingency of their future relationship, and legislators cannot predict all the circumstances under which their bills will be applied, we cannot *ex ante* specify "if-then" rules that provably direct good AI behavior. Legal theory and practice offer arrays of tools to address these problems. For instance, legal standards allow humans to develop shared understandings and adapt them to novel situations, i.e., to generalize expectations regarding actions taken to unspecified states of the world. In contrast to more prosaic uses of the law (e.g., as a deterrent of bad behavior), leveraged as an expression of *how* humans communicate their goals, and *what* society values, *Law Informs Code*." (NAY, 2023: pàg. 309)

> We are not aiming for AI to have the legitimacy to make or enforce law. The most ambitious goal of *Law Informing Code* is to computationally encode and embed the generalizability of existing legal concepts and standards into AI. Setting new legal precedent (which, broadly defined, includes proposing and enacting legislation, promulgating agency rules, publishing judicial opinion, systematically enforcing law, and more) should be exclusively reserved for the democratic governmental systems expressing uniquely *human* values. Humans should always be the engine of lawmaking. The positive implications (for our approach) of this normative stance are that the resulting law encapsulates human views. (Nay, ibid. p. 326)

Ambiguity is assumed as consubstantial. NAY's proposal is to deal with it beforehand. For example, the so-called "legal gaps" (state-action pairs without a value) are often treated with the invocation of standards such as "material" or "reasonable". NAY (ibid. p. 366) proposes to use them as "modular building blocks (pre-trained models) in AI systems". It is therefore expert knowledge, such as that of CELSIUS or TUBERO, which can inform the way of regulating.

Another proposal along the same lines is *Constitutional AI*. CLAUDE, the proposed chatbox, is interesting because it has been trained with ethical principles aimed at eliminating ambiguities about the morality of the response. The main objectives are (i) to scale supervision, (ii) to eliminate evasive responses, (iii) to improve transparency, (iv) and to reduce iteration time by obviating the need to collect human comments. The AI assistant CLAUDE can (i) engage with harmful queries by explaining their objections to them, (ii) and leverage chain-of-mind style reasoning using a dual supervised learning (ML) process and reinforcement learning (RL). First, sampling, self-criticism, and tuning occur from a starting mode (SL). Then, in the reinforcement learning phase, a model is used to evaluate which samples are better, producing a preference model using RL of AI Feedback (RLAIF). Thus:

> We will be experimenting with an extreme form of scaled supervision, which we refer to as Constitutional AI (CAI). The idea is that human supervision will come entirely from a set of principles that should govern AI behavior, along with a small number of examples used for few- shot prompting. Together these principles form the constitution. (BAI *et al.* 2022)

With the development of systems based on GPT-4 pre-training, or others that encode values, such as CLAUDE (from Anthropic), AI has indeed taken a step forward in modelling human intelligence. However, the recent systematic comparison of six chatbots—ChatGPT-3.5, ChatGPT-4, Bard, Bing Chatbot, Claude2, Aria—leads to the conclusion that we are still far from equal (LOZIC *et al.*, 2023). AI certainly shows limited potential in terms of originality of contributions. It is better to understand it as a set of instruments of a cognitive nature, but even from this point of view its potential for transformation is enormous; *disruptive*, to term it with CHRISTENSEN (2013).

## 11. Regenerate

There is a problem here that should be clarified. Both jurists or legal researchers and engineers have a vision of regulation that comes from the perspective inherited in the 19th century, be it legal realism in the systems of the *Common Law* or (inclusive or exclusive) positivism in the *Civil*

*Law*. Thus, they consider that there is an established process of creation, interpretation and application of law that comes from a set of codes or laws written and promulgated by a power or authority that operates to regulate human behaviour. The ambiguity operates either in the normative language, or in the relationship between this language and the conduct of the subjects. We have seen this in the theory of legal interpretation by WRÓBLEWSKI and DASCAL. They assumed that the triad (i) *law-in-natural language*, (ii) *law-in-language* (formal or semi-formal), and (iii) *law-in-behaviour* was enough (patterns of behaviour based on social rules, for example).

However, in a symbiotic or hybrid reality between humans and machines, this triad should take into account at least: (i) *agents* that are computer programs with various degrees of autonomy (Multi-Agent Systems, MAS), (ii) *cyber-physical systems*, in which information processes are already imbued within the devices that are used daily (from printers to automated cooking machines), (iii) and *online institutions*, capable of convening a set of users ranging from a few dozen to literally hundreds of millions (as is the case with Google or Twitter).

These systems operate through digital platforms that can control their information flows. Thus, a minimal ontology of subjects / objects / actions / relationships is represented as digital 'entities'. We could therefore add a fourth element to the triad: (iv) *law-in-AI enabled transactions* The nuance matters because we are not dealing with classic legal structures, but with new instruments that expand the field of regulation and modify its nature. *They are regulatory systems rather than legal systems*. Or, to put it in a more precise way, regulatory systems capable of generating *legal ecosystems*.

From this perspective, *augmented reality* (AR) *digital twins* (DT) and what has been called *metaverse* represent a challenge that we are just now beginning to face.[47] And here is a new phenomenon of ambiguity, where we must carefully distinguish between the physical, real subject and the digital twin, the digital entity that represents us in the digital world as citizens.

The latest version of the *GARTNER Hype Cycle for Digital Government Services* defines it as follows:

> A digital twin of a citizen (DToC) is a technology-enabled proxy that mirrors the state of a person. National, state and local governments use DToC to support citizen services such as health or safety management. Its elements are the model, data, a unique one-to-one association and the ability to monitor it. It integrates data into the DToC from siloed sources such as health records, credit scores, phone logs, criminal records, customer records, and sensors such as cameras. (GARTNER (2023: p. 43)

We now come to the notion of *legal ecosystem*. There are two ways to understand this notion, which is starting to be used in the world of law operators dedicated to Web services. The first is to refer to the complex of agents, regulatory systems, practices, and institutions that define what we understand by 'law' (national, international, global).[48] The second is to refer to platform-driven economy, and in this case we find real-time regulation from which an intelligent or *smart ecosystem* emerges.[49]

---

[47] ''Metaverse' is not a technical, but a commercial one. It indicates a virtual reality space where users interact with and within a digitally generated environment. See about the expected impact in various areas (from industry to government), GARTNER (2022a, 2022b).

I will put the example of the EU Project OPTIMAI, a European project to build an intelligent platform for the industrial sector (*smart manufacturing*).[50] The validity and traceability of transactions are ensured stemming from: (i) the decentralization produced by an authorized blockchain with an access control layer; (ii) the use of Ethereum with Proof of Authority (PoA) consensus mechanisms; (iii) executed smart contracts; (iv) the middleware that controls access and provides data to the block chain (MARGETIS *et al.*, 2022). Hence:

> The aim is to store and retrieve data collected on the production line using Ethereum-based private blockchains, each of which is dedicated to a specific factory. […] each factory's products go through quality control, where sensors and AI methods analyze the objects to detect defects. If a defect is detected, the data is sent to the middleware, which sends a JSON file to the appropriate endpoint IP of the Blockchain API. The Blockchain API then sends the data from the JSON file to the smart contract, which is deployed on the private Ethereum blockchain. We utilized Geth, from "Go Ethereum" a tool used by Ethereum, to set up the Ethereum nodes and created three privates Ethereum blockchains, one for each factory, as each factory's data is private and cannot be shared with others. (MITSIAKI *et al.* 2023: 1176)

Figure 5 does not properly constitute a figure, nor is it any technical representation of how the modules that make up the platform work. It is a schematic view. But it is enough to illustrate the points I have pointed out to understand the components of intelligent ecosystem.

---

[48] "A *legal ecosystem* can be defined as a complex and dynamic system that includes multiple levels of governance, ranging from local to national and international, and involving a wide range of actors, including lawmakers, judges, lawyers, law enforcement officials, civil society organizations, companies, corporations, and ordinary consumers and citizens". (CASANOVAS, HASHMI, DE KOKER 2024). *Vid*., specifically on OPTIMAI, CASANOVAS (2024).

[49] "A *smart legal ecosystem* (SLE) is (partially) embedded into cyber-physical systems that function in an intelligent environment encompassing the features of the IoT 4.0 and 5.0 (especially ethics and law), to achieve legal compliance in real time." (CASANOVAS, HASHMI, DE KOKER 2024, *ibid*.)

[50] OPTIMAI. *Optimizing Manufacturing Processes through Virtualisation and Artificial Intelligence.* H2020-(IA) DT-FOF-11-2020. https://optimai.eu/

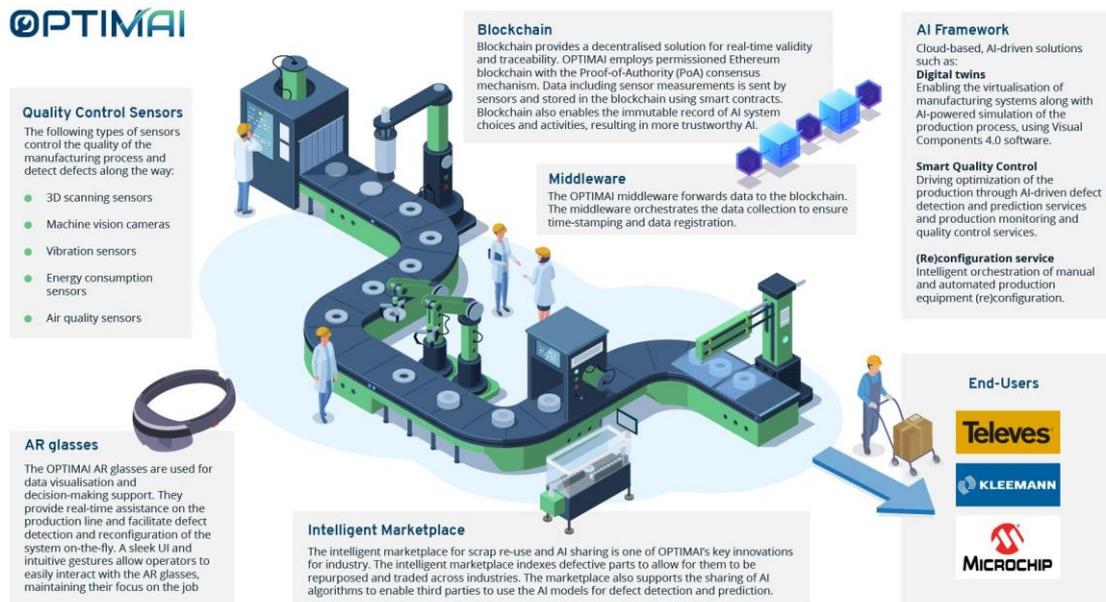

**Fig. 5.** Synthetic view of OPTIMAI components. Source: https://optimai.eu/

There are quite a few advantages from this perspective. In addition to increasing visibility and traceability, it generates *trust* among all partners and users:

> Smart contracts come handy when there is a need for stringent evaluation against a set of conditions and constraints. Examples for smart contracts can be used to develop a supplier evaluation system, onboarding supplier in case of immediate requirement, maintaining cold chain requirements for healthcare and food supply chains, and alerting the corresponding team in case of any violation, tracking the products throughout their lifecycle etc. (KOTHA AND SONY, 2023: p. 31)

Nevertheless, ambiguity turns out to be a feature here as well that is not, by any means, residual. Smart contracts are agreements from a technical point of view, but they do not constitute legal contracts by themselves (DE FILIPPI AND WRIGHT, 2014). In fact, crypto-currencies and block chains are being used for illegal reasons also in the so-called *dark web* (DEMETIS, 2023) In addition, the ambiguity and uncertainty of the regulations in each country is one of the issues pointed out as obstacles (WANG *et al.* 2023; KOTHA AND SONY, 2023). The modelling of legal compliance and the way to describe and measure it is one of the remaining problems that still must be solved in a minimally satisfactory way.

## 12. Comply

My position on relational law does not entail that it is impossible to model ambiguity (as asserted by BIRHANE, 2021). We have already seen that there have been successful attempts, albeit with strong assumptions. The field of engineering aimed at regulatory systems to carry out audits and legal compliance has had a spectacular development in the last twenty years, since the scandals of Enron, Arthur Andersen, and Adelphia Communications Corporation, among others, gave rise to the Sarbanes–Oxley Act of 2002, which in turn encouraged the creation of compliance officers in virtually all corporations. Ever since, automated internal audit languages and processes flourished. These languages apply formal rules to model, segment and monitor compliance processes. This does not mean that interpretation is not relevant. On the contrary, it is one of the tasks entrusted to officers.[51] Ambiguity and ambivalence (the internal position of officers in the organization) go alike, because compliance officers (i) act as a bridge between external and internal investigations and audits; (ii) they are assigned tasks of regulation, training and control.[52]

Compliance is not obedience, but *conformance*. Despite the stipulative definition of many dictionaries[53], 'compliance' in this sense has more to do with the orderly and controlled monitoring of processes than directly with a system of sanctions that would operate from a vertical authority or power. These exist and the violation of the rules is taken into account in order to apply a system of sanctions, but the dynamic of following the processes correctly is more important than pursuing those who break them by default. They are more preventive than punitive, and that is why they have been set up, to avoid fines and lawsuits. Thus, compliance officers can be called "double agents" as they are "literally double acting on two different levels: in the interests of the company that hires them, and in the interests of the market as a public good" (LENGLET 2012: 66).

It is this intermediate position that makes them interesting, because the automation of processes can be carried out through a multitude of business languages. But there are functions, such as strategic advice on market policies or lobbying functions, that fall outside the scope of automatic regulation. Corporate strategies also matter because they integrate compliance or non-compliance as a form of operational behaviour. For example, in finance, the phenomenon of excessive compliance or *over-compliance* usually occurs in relation to the denial of services to clients as risk reduction (*de-risking*).[54] In a conservative strategy, drivers of over-compliance impact on decisions of banks to terminate relationships with customers and counterparts.

---

[51] "Compliance is one of the functions taking an active part in this translation, building on its specific position within the organization. Compliance officers act as employees contributing to the collective construction of profitability, while at the same time ensuring that the resulting practices stay within the paths delineated by law. Bridging representations and communities from the inside, compliance officers embody an agency relationship between management and employees, the company and the market, the inside and the outside, the tacit and the explicit within the organization. In enacting internal norms and performing controls, advising and training employees, compliance officers are actors in the regulatory architecture now in place in the majority of Western-style economies. As such, they are said to manage the 'reputational' or 'image' risk arising from non-compliant practices, thereby encountering the 'morphing meanings of risk." LENGLET (2012: 60).
[52] "As members of one control function among the long list of market actors usually remaining backstage, compliance officers contribute to internal regulation of the market by managing the *ambiguity* arising from the encounter between texts and contexts. I argue that this process is possible because of the *ambivalent* position they hold within the organization" LENGLET, op. cit. p. 61.
[53] For instance, the *Cambridge Dictionary* definition: "the act of obeying a law or rule, especially one that controls a particular industry or type of work".

There are very few papers yet on ambiguity and rule following. It is estimated that just over half of the requirements of the so-called *Automated Compliance Checking* (ACC) are checked manually, mainly due to the poor semantic determination of the requirements, which need to be interpreted. In a taxonomy just in the process of publication, ZHANG *et al*. (2023) have verified what we knew from experience: some clauses are ambiguous due to opaque linguistic usages, tacit knowledge, and specific reasons of regulators; while others are intentionally ambiguous to adapt them to a changing context and circumstances that are left to the auditors' discretion (ZHANG *et al*. *ibid*., 2023).

This is a well-known procedure in law. Drafting laws in natural language means modulating the ambiguity of the wording so that the application of the law can be graduated depending on the interpretation that is considered most appropriate. There is an open door to the interpretative plurality of the rules precisely because precision is achieved at the moment of making them effective. At the general level of the formulation of regulatory provisions, their meaning, not their sense, should be clear. Their sense is determined in specific situations and applications, and that is why interpretation is relevant. Even when the wording is transparent, this door is left open. It is one of the problems that must be raised and solved in the modelling of rules for their execution, as we will see in the following sections. I will give an example.

In the *CRC Data to Decisions* program, a joint endeavour of the Australian Government with universities and research centres[55], the Australian Criminal Intelligence Commission raised the problem caused by having to issue criminal certificates. To access a position or to access a job, you must present them; *ex officio* if it is in the public service, or at request of the employer. The issue is whether there is an obligation to disclose the convictions suffered, and until when. After a few years, it is considered that the sentence has already been served or has expired, and therefore there is no longer an obligation to make it public.[56] This is done to protect the normal development of people's social life. Each year more than five million requests are received that must be resolved and have a long waiting time. Would there be a way to ease the process safely and satisfactorily?

---

[54] "Over-compliance in a rule-based context means going well beyond what the law requires or staying very clearly well within its boundaries" (de Koker and Casanovas, 2024, in press).
[55] https://www.d2dcrc.com.au/
[56] "A 'spent conviction' is a conviction of a Commonwealth, Territory, State or foreign offence that satisfies all of the following conditions: (i) it is 10 years since the date of the conviction (or 5 years for juvenile offenders); AND (ii) the individual was not sentenced to imprisonment or was not sentenced to imprisonment for more than 30 months; (iii) AND the individual has not re-offended during the 10 years (5 years for juvenile offenders) waiting period; (iv) AND a statutory or prescribed exclusion does not apply. (A full list of exclusions is available from the Office of the Australian Information Commissioner)."
https://www.afp.gov.au/our-services/national-police-checks/spent-convictions-laws-police-checks
https://www.nationalcrimecheck.com.au/faqs/spent-convictions-schemes-general-information

The *Spent Convictions Scheme* is clearly regulated in the law [*Crimes Act 1914 (Cth): ss 85ZV and 85ZW*]. How can it be modelled? First, the rules had to be extracted from the provisions of the text, with the specification of which parts were ambiguous, which required interpretation, and to what extent it was possible to provide automatic solutions. The very expression "literal interpretation"—so common in legal doctrine—is ambiguous. It turned out that even in a clean draft like this one *there were nineteen issues on which decisions had to be made*. JEFF BARNES (2019) distributed them according to the potential issues, that is, depending on the arc of possible difficulties when moving from the abstract meaning to the specific sense of the text in its practical implementation. He distinguished between (i) those that had been created intentionally, (ii) unintentionally, (iii) and those that could give rise to different interpretations based on expert legal knowledge—i.e., jurists who knew the techniques included in the *Acts Interpretation Act 1901 (Cth) ('AIA')*.[57]

The analysis is interesting because it shows that ambiguous expressions constitute only a part of the problem and require comparatively simpler decisions to make. What does "Commonwealth law" mean in the scheme? What does the expression include? The text mentions "instruments". Are they necessarily only "legislative instruments"? As BARNES (2019, p. 12) asserts: "A judgment will need to be made on a case-by-case basis as to whether an instrument, that is not described as a rule, regulation or by-law, partakes of a legislative character."

Another example of ambiguity is the meaning of "or otherwise" in the following paragraph:

> (2) Section 85ZW(a)
> Subject to Division 6, but despite any other Commonwealth law, or any State law or Territory law, where, under section 85ZV, it is lawful for a person not to disclose, in particular circumstances, or for a particular purpose, the fact that he or she was charged with, or convicted of, an offence:
>
> (a) it is lawful for the person to claim, in those circumstances, or for that purpose, *on oath or otherwise*, that he or she was not charged with, or convicted of, the offence; and ...
> (b) *anyone else who knows*, or *could reasonably be expected to know*, that section 85ZV applies to the person in relation to the offence shall not…

The interpretation is relevant because it is necessary to decide whether it is accepted as 'legal' for a person to declare about his past with an act similar to an oath, or if another means is also acceptable, e.g. a mere statement. The problem continues with the meaning of "anyone else who knows, or could reasonably be expected to know".

Only after reaching a decision on these issues, GUIDO GOVERNATORI was able to build the more than two hundred rules that were finally modelled with the *Turnip* editor to carry out the proof of concept (GOVERNATORI *et al*., 2020). But even so, we could verify that the necessary interoperability did not yet exist between all the applicable provisions to fill the concepts with meaning automatically.

---

[57] Accessible at https://zenodo.org/records/3271515 .

**13. Execute**

Rules can be followed, in many ways, depending on the instruction and nature of the rule. But rules, in computing, are *executed*. Again, the term is already ambiguous in itself, because it has a wide range of meanings that goes from the different degrees of compliance and management to all the degrees of implementation of the actions regulated by modal operators (permissions, obligations, prohibitions, etc.). *Execution* has connotations (i) of strict interpretation of the rule, (ii) but also of the seriousness of the acts that are its effect or consequence. One 'executes' a loan, an eviction, or even a person. *Executioner*, in English, means executioner (hangman, headman). In fact, the etymology of the word comes from the French-English of the end of the 13th century, *execucioun*, 'to carry out', 'to finish', and from the Latin *exsequor*, 'to follow up to the end', 'to continue', where the term 'exequy' comes from as well. The suffix *ex* (head out) indicates 'materialisation', 'realisation', 'bring to effect'. To produce effects, therefore, that change the environment, the 'outside' world.

There is a mechanical part, of necessity (*ananké*), so that the effect follows necessarily from the conditions. Sometimes, in computer languages, the relationship is not only logical, but causal, in the sense that, as information, it is processed from the *activation* of the rule and becomes, triggers, a result that it has an impact on the context.[58] The rules, understood as a kind of algorithms, work like this. I think this is why the late WITTGENSTEIN focused so much on it in the second part of his work: "Following a rule" has to do with the *actions* it can give rise to, inside and outside the semantics of the verb or the name to which it refers. From this point of view, the '*nomotropic*' space that A. G. CONTE (2016) was talking about following the Viennese philosopher, i.e. *reacting* to a rule, is literally not computable, since it involves a number of unknown variables. 'Unambiguity' cannot exist, "since there is always a gulf between an order and its execution".[59]

ROSS's example, "Either slip the letter into the letter-box or burn it!" (see note n. 3) does not exclude other actions that can be done with the letter. Simple ones, like framing or tear up it; or more complex, such as publishing it or turning it into a symbol of resistance or blasphemy. No defined, not yet computable future actions, as A.N. WHITEHEAD would write in 1911 and Stuart Russell is reminding wisely, are the basis of culture.[60]

---

[58] See Note 3, we could add that WITTGENSTEIN notion of rule and conditional lies behind PRIEST's attempt to model paradoxes.

[59] Cf. WITTGENSTEIN (1953, 1978, I # 433, p. 188ᵉ) "When we give an order, it can look as if the ultimate thing sought by the order had to remain unexpressed, as *there is always a gulf between an order and its execution* [my emphasis]".

[60] "Civilization advances by extending the number of important operations which we can perform without thinking about them" (A.N. WHITEHEAD, cited by RUSSELL, 2023, p. 88).

My thesis is that, in computing, we also find ourselves as explorers in the unknown country that WITTGENSTEIN imagined when talking about orders and rules in *Philosophical Investigations* (1953).[61] By the very fact of being so, natural languages are instrumental, necessarily fluid, open, usable and reusable, to perform an indefinite number of actions in changing ontological frameworks as well.

The practice of *moral* decisions in specific situations and problems cannot be derived or inferred from theories that are situated in a more abstract dimension, level, and linguistic framework of reflection. There is not, because there cannot be, an ethical theory from which the solutions are linearly deduced. There is *applied ethics* instead, focusing on the variety of problems that arise in different implementation fields according to also different knowledge (medicine, law, or engineering) (CAMPS, 2013).

Although it may seem otherwise, computational languages, both symbolic and generative AI, maintain this pragmatic property, but in a different way. They reproduce a referential and meaning ambiguity that they try to close with a sense that emerges from the very moment in which it is formalized. We say that a musical work is *executed* (and not a conversation or a literary work) because we assume that the performer follows a previous plan, i.e. the work that the composer has captured on the staff. We also say that computer programs are executed for the same reason. Likewise, the rules of a computer program are executed according to its design.

But this should be nuanced because, in fact, what they do is generate a computational field that can lead to what we have called *dynamic ecosystems* (in our case, ethical and legal), also variables, that can evolve intelligently. The issue is different if the question is (i) how the pragmatic dimension of natural language can be coded, i.e., how can their functions be 'captured'?, (ii) or how can they be managed, i.e., do AI languages also function as natural language games, as "activities" or "forms of life"

I think that both in the case of symbolic AI and in that of generative AI, *applied to regulation*, the answer should go one step further than saying that ambiguity is ineliminable. What is ineliminable is the *interface between language and context*. If ambiguity is a linguistic property that deals with the indeterminacy of the reference, soon the linguistic context must confront the environment in non-virtual situations or scenes, and these can intervene in the modification of the rules in an uncontrolled or unforeseen manner.

---

[61] "Suppose you came as an explorer into an unknown country with a language quite strange to you. In what circumstances would you say that people there gave orders, understood them, obeyed them, rebelled against them, and so on? The common behaviour of mankind is the system of reference by means of which we interpret an unknown language." (WITTGENSTEIN, 1953, 1978, I #206, p. 82ᵉ).

## 14. Contextualise

In design sciences research (DSR), the term 'context' is used to refer to the environment or ambiance in which something exists, i.e., the aspects of the environment that are relevant to explaining a phenomenon, the environment surrounding an artifact, or the source of the requirements with which an artifact must be evaluated (HERWIX AND ZUR HEIDEN, 2022). From a cognitive perspective, the view of context is pragmatic, situational, and *distal* (depending on the perspectives of each of the actors involved). According to these authors (*ibid*.), nine dimensions should be distinguished: (i) *function* (awareness, improvement, integration); (ii) *focus* (problem, strategy, solution); (iii) *scope* (local, domain, global, universal); (iv) *reference field* (practice, academic); (v) *perspective* (technical, mixed, social); (vi) *orientation towards facts* (factual/counterfactual); (vii) *historical orientation* (past, present, future, invariant); (viii) *awareness of time* (dynamic, static), (ix) *control* (natural/artificial) (ibid.).

The execution of rules depends on their structure. To activate them in a computational language, we should identify: (i) the *agents* (subjects or stakeholders being involved); (ii) the *preconditions*; (iii) the *conditions;* (iv) the *triggers*; (v) the *connector*; and (vi) the *regulatory effects* of permission, obligation, or prohibition that are intended to occur (type of norm). We adopt the formulation of MUSTAFA HASHMI (2015), although other descriptions of rules could have been considered as well (RIVERET *et al*. 2024).

(1) Structure IF (*antecedent*) THEN (*consequence*):

If $(A_1; A_2; \ldots A_k)\ldots$ THEN $\{B\}$
where $A_1; A_2; \ldots A_k$ are the antecedents (conditions) of the rule and B refers to the legal effects of the implementation of the rule when conditions are met

(2) We add the types of rules and the properties of the process to the structure IF… THEN of equation 1;

*IF* (Conditions of the rule)

   *THEN*

   *(Type of norm)* $|^{at\ (element)}$ {Legal effects$_i$,…, Legal effects$_n$}

The symbol $|^{at\ (element)}$ refers to the aspect of process to which a rule is applicable (or to the produced legal effects), and the subindex i=1… n is a natural number (*n-th*). The term "legal effects" refers to the desired consequences resulting from the effects of the applicable conditions of the rule. Let's take as an example the rule [1] formulated earlier in Section 5 (*Regulate*):

[1] If a driver (S1) wants to turn right at an intersection with a red traffic light, she (should, must, has to, necessitates):
  (i) Violate the rule prohibiting passage through red lights;

(ii) Wait until vehicles approaching from the right (S2) and from the front (S3) have passed;
(iii) Execute the right turn before vehicles approaching from the left (S4) and from the front (S5) start moving when they have a green signal.

In this case, the *preconditions* are that the driver has the vehicle ready to turn; the *conditions* are being about to turn right and the traffic light being red (i.e. the norm has been violated); the *trigger* is that vehicles from the right and from the front have finished passing; the *connector* is a semantic implication; and the *effect* is the resulting permission to make the right turn. Note that there is also a causal effect: actually, making the right turn, which is both permitted and caused by the fact that the material conditions of the rule have been met. Regarding the prohibition of crossing a red light, it could be said that it remains 'clipped' while the vehicle is positioned to make the turn. The point I would like to emphasize is that to resolve contextual ambiguity, we must also represent the context at the same time as we construct and formalize the rule in some computational language (for example, in temporal event calculus).

Regarding the *connector*, we note that this approach is compatible with the three dimensions of the $\mathcal{WITT}$ model for *online institutions* (OI) by Noriega *et al.* (2023).[62]

The same rule can be understood from these three different perspectives: *causal, logical, and technological*. A more detailed comparison with the metamodels developed for multi-agent systems (MAS) in OI would be worthwhile. I won't do it now. I will limit myself to observing that (i) twenty years earlier, the notion of "*interlocking norms*"[63] already provided that the regulatory link could "trigger" obligations based on regulatory non-compliance or the violation of rules by artificial agents; (ii) the notion of "context" has received preferential attention as the notion of "electronic institution'" (and now that of 'online institution'-OI) have been conceived operating already in an open space where their social interaction with humans takes place; (iii) therefore, the violation of norms both by humans and by intelligent agents also allow the modelling of *contrary-to-duty conditions* (vid. above, Section 5 *Regulate*).

But this position is also based on an *analogous* perspective regarding virtual institutions, which are supposed to symmetrically reflect the structure of conventional institutions, i.e. they are understood as *analogues* of offline institutions.[64] The subsequent link of the compliance of the agents to the enforcement and to mechanisms aimed at correcting the violation point to a *normative conception of the rules* that, for what has been explained, I do not quite share.[65] The *regimented* rules of production, i.e. the conception of rules representable in finite state machines, have limits.[66] I prefer to conceive the rules in a singular way, in a less unitary and systematic way, so that they can be applied to complex variable situations in a multitude of different contexts.

---

[62] Cf. NORIEGA *et al.* (2023). "$\mathcal{W}$ corresponds to the fragment of the real world that is relevant for the activity that is performed within the OI, $\mathcal{I}$ is an abstract representation of $\mathcal{W}$ that establishes the 'rules of the game' and thus provides the specification of how the OI is meant to operate, and $\mathcal{T}$ consists of the information technology that implements and supports the OI." I thank PABLO NORIEGA for this observation.
[63] Cf. LÓPEZ AND LUCK (2003). "The norms of a system are not isolated each other; sometimes, the compliance with some of them is a condition to trigger (or activate) other norms. That is, there are norms that prescribe how some agents must behave in situations in which other agents either comply with a norm or do not comply with it."
[64] Cf. D'INVERNO *et al.* (2012: 38): "Just as conventional institutions are organisational structures for coordinating the activities of multiple interacting individuals, electronic institutions provide a computational analogue for coordinating the activities of multiple interacting software agents".

The context can be represented from the nine dimensions we have mentioned. But, even so, it must be distinguished from *patterns of behaviour* and from the *specific environment* in which the rule can operate. Environments constitute specific scenarios, with their own exogenous variables (number of drivers and tracks, track width, etc.).

The Fig. 6 is a recent attempt to combine all the elements needed for the extraction of rules from provisions or normative texts expressed in a natural language, also due to MUSTAFA HASHMI'S work. The contextual dimension is what allows the content and operation of the rule to be defined in the pragmatic dimension and level of specific scenarios, i.e. the *micro level* of the construction of the ecosystem based on the activation of rules and the use made by the agents.

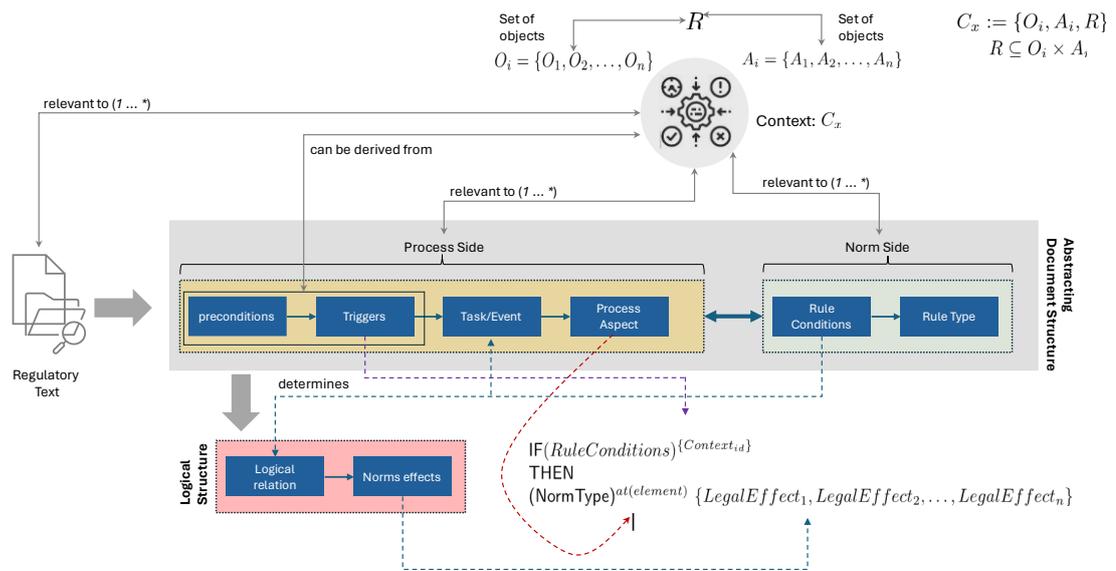

**Fig. 6.** Architecture of the extraction of rules from regulatory provisions, with the incorporated context. Source: HASHMI (2015), HASHMI AND CASANOVAS (2024, work in progress).

---

[65] It comes from the version of '*norm*' (Norm, in German) that the Neo-Kantian positivist jurists of the second and third schools of German public law elaborated (last part of the 19th century, first third of the 20th century). It passed to MAX WEBER via GEORG JELLINEK and HANS KELSEN, and from there to European and American sociology of the 20[th] century. The uncritical reception of the dichotomy—'*social norms*' and '*legal rules*'—has given rise to a 'normative' perspective in MAS which has been useful but which has also created a framework which has been generally adopted as a guideline, but which has no reason to be the only one. For example: "Societies are regulated by norms and, consequently, autonomous agents that want to be part of them must be able to reason about norms." (LÓPEZ AND LUCK, 2003). I appreciate the relevant comments and discussions with Pablo Noriega on this point.

[66] I thank the relevant observations and discussions held with PABLO NORIEGA on this issue.

In a complex case, such as OPTIMAI (see Fig. 5, Section 12, *Regenerate*), the smart ecosystem can be *re-expressed* through rules that operate on the flow of information that is generated from the smart contracts and the private blockchain (Ethereum). This creates a context that includes both human and artificial agents, and that can give rise to a multiplicity of different scenarios, each with features (attributes) and specificities at the micro-level in which they are located. In OPTIMAI we are dealing with manufactures with different products —microchips, antennas and elevators— which are capable of carrying out quality control and saving human and material resources from a self-control generated and supervised semi-automatically, with the incorporation of ethical principles (values) and legal norms (rules) inside and outside the regulatory system they adopt.

The regulation operates in real time in augmented reality, simulations and in the human/machine interface. I don't think this can be properly described by traditional concepts such as "law-making" and "law-application (or enforcement)". Cyber-physical systems, robots and *mobots*, are articulated in systems that include human decisions, but which can normally function automatically. They produce data and metadata being (i) recorded through *sensors* and physically activated through *actuators*, (ii) analysed through AI techniques (deep learning) [67], (iii) and controlled through simulations and flows of encrypted connection information with the platform.

In order for this to happen, it is true that paradoxes must be avoided, systems must not present inconsistencies, and the semantics of the "open texture" of the language presented by common sense concepts must be closed as far as possible. Legal theorists who are experts in technology and AI have kept insisting on the creation of formal legal systems of this kind (VAN DER TORRE AND TAN, 1999, 2017a, 2017b; SARTOR, 2005, 2009). I partially agree. My position does not involve the conception of abstract and closed normative systems, but the creation of instruments that operate at the most specific level of implementation and that, therefore, can be fragmented, manipulated, and used as institutions by the users. They have both an instrumental and constitutive character.

We have just seen that the open texture of natural language is not limited to the concepts of positive law that must be defined to appropriately used —such as 'property', 'crime', 'norm' and 'penalty'. Methodological concepts such as 'rule', 'violation', 'condition', 'inference', etc. should also be specified in their epistemic and practical dimension to produce regulatory systems that end up generating ecosystems. To put it another way, there is a way forward and another way back from natural languages and general knowledge of common sense meanings to technical and scientific concepts. Law is often as occlusive as it is transparent. But so does machine language, and this applies to both symbolic (rule-based) and generative AI. Sooner or later, in the initial definitions and in the implementation and refinement of systems, natural language and its accumulation of knowledge operates through artificial languages and systems built.

---

[67] Per a ser precisos, OPTIMAI usa la tècnica denominada *focused concept miner* (FCM), un algorisme interpretable de mineria de dades que extreu conceptes d'alt nivell de dades textuals. Cf. Lee *et al*. (2018).

**15. Re-express**

We have come to the end, and it is time to wrap up. In the second part of the article I have described the *double-loop* of the language of law, and I have criticized the occlusion that occurs in the attempt to comprehensively 'elucidate' the perspectives, concepts and theories considered 'ambiguous'. On the contrary, I have argued that accepting a certain degree of ambiguity is the first step to achieving an analytically clear outcome and being able to face problems in digital environments.

I am not alone in this. During the 19th century and until the first Great War, the so-called 'legal certainty' (referred to 'security' in the Civil Law tradition: *Rechtssicherheit, seguridad jurídica, sécurité juridique*…) dominated the panorama of reflections on the functions of law. Then the door was opened to reflections of a hermeneutic nature, where human rights and, above all, the Constitutions, achieved a preeminent role in the definition of the legal. This led to interpretative and argumentative theories, which have been predominant until today. It is not a matter of underestimating anything now, but of taking advantage of everything possible because it is not the law, but the world in which it is applied that has changed radically.

A partir de la crisi econòmica del 2008-2012, es va obrir pas una posició en innovació i estratègies de mercat coneguda pel seu acrònim (VUCA). Contempla el món digital en els mercats i en les organitzacions tal i com es presenta: (i) *volàtil* ; (ii) *inestructurat*; (iii) *complex*; (iv) i *ambigu*. És un món on l'agilitat en la reacció i el joc ho són tot:

Stemming from the economic crisis of 2008-2012, actually a bit before, at the beginning of the Millenium, it a position in innovation, leadership, and market strategies known by its acronym (VUCA) flourished. It contemplates the digital world in markets and organizations as a competitive space: (i) *volatile*; (ii) *unstructured*; (iii) *complex*; (iv) and *ambiguous*. It is a world where agility in reaction, resilience, strategy, and game is deemed to be the key to the problem:

> By *definition,* in a VUCA world, if you're not confused, you're not paying attention. Confusion is part of the game. And actually, being frightened is part of the game, too. But you cannot *stay* frightened, or you will freeze and lose the game. […]. The ugly truth of the VUCA world is that clarity gets rewarded, even if it's wrong, because there is such a need to cut through the confusion. A lot of people can't live with the current level of confusion, so they find simplistic solutions attractive. And of course that can be very dangerous. What you need is to be clear and simple without being simplistic. And then finally, *ambiguity yields to agility* [my emphasis]. In the VUCA world, you have to be an athlete to thrive. (JOHANSEN AND EUCHNER, 2013: p. 10).

I would now like to go back to the beginning. In this article I have also been able to distinguish some nuances referring to the language, organization and technology of law. I have distinguished between *meaning* and *sense* and, as I promised, I have ordered my discourse by pivoting the different sections on ambiguity. Even if provisionally, it is time to draw some more general conclusions, so that the discourse *makes sense*. I am doing it from my on field, law and technology, without any further pretensions.

1. Ambiguity is a feature of natural languages that is inescapable from any theoretical perspective we adopt in the use of formal or semi-formal languages. What it reveals is the problem of the connection of language with actions and the context in which they occur.

2. It is not a language defect, a negative trait or an obstacle that needs to be overcome. Quite the opposite: it is a way of making it flexible, adapting and using the cognitive and communicative resources we have to solve problems and regulate conflicts in complex situations, environments and scenarios.

3. From this perspective, we can deal with ambiguity with a host of techniques to reduce its effects and minimize its possible risks in specific contexts, situations and scenarios. But we can also consider it as an advantage, an important point of departure that should be considered in the sciences of design (think of its abductive aspect, for example).

4. We should therefore distinguish not only the different levels of autonomy that are present in multi-agent systems (MAS) or in online institutions (OI), but also the different ways of regulating the degree of maturity held by different computing technologies—block chain, semantic web, AI (LLM, FM, NLP, DML).

5. Also, basically, we should learn to use concepts well to take advantage of their different uses in regulation and have the capability to figure out some new ones. The perspective of *intelligent, emerging and sustainable legal ecosystems* in real time, I think is more than a theoretical proposal. It is rather a reality that has been developing in practically all areas of regulation —from industry to health and administration—and that we would do well to consider in all its complexity.

6. Therefore, measuring from an empirical stance the degree of regulatory and legal compliance seems the way to go. In a "platform economy", where everything has become "as a service" (i.e. mobility "as a service", or legal platforms "as a service"), we should be attentive to what the results of these services are. Especially, how the knowledge shared with users (as citizens and consumers) is distributed and intensified (or not). [68]

7. Thus, from the point of view of legal technique, we should add to the classic theory of interpretation the specificity of the systems and techniques that come from the sciences of design. At the same time that (i) *law-making*, (ii) *law-applying* (iii) and *law-describing*, we should consider (iv) *law-processing*, i.e. *law-in-transactions through AI*, responding to a hybrid or symbiotic reality between humans and machines (H/I) that presents different features than those we knew until now.

---

[68]There have already been acid criticisms of the extractive power of platforms (CRAWFORD, 2021), and STUART RUSSELL 's book (with the *Afterword* of 2023) makes it clear what can happen if the idea of a general intelligence that can control the intelligence of its own creators is accepted uncritically.

I have tried to show that linguistic and cognitive pragmatics can constitute a good epistemological base point in order not to oversimplify and thin out a reality that, contrary to what is usually assumed, is thicker than before. This *hybrid*, *symbiotic* reality, between humans and machines, is not lighter, no matter how much the information systems operate digitally. Humans are always in the middle, be they mediators or meddlers in HRI.

Thus, we return to the expressive function of law. I understand law as a set of regulatory techniques, among others. At this stage, I do not think that the analytical distinction between the *hylomorphic* and the *expressive* function of norms is useful anymore. Rather, we should leave them aside in the construction of intelligent designs that are characterized by their ability to integrate (i) a plurality of theoretical positions (ii) and a multitude of different technologies.

RUSSELL (2023) has pointed out two points that are worth considering. The first is that robots can abide by rules and follow rules — that is, follow "the letter of the law." — but what is difficult for them to do is to follow its "spirit" because they cannot understand. This would involve activating knowledge of common sense that they simply lack (*ibid.*: 203). The second is the difficulty of implementing effective protections—for example, the privacy defended by the GDPR or the AI Act risk manage provisions, because the development of more effective information services inevitably also implies a greater capacity for control and an increase in risks. This is one of the main dilemmas we pointed out recently about the governance of AI and its use as a regulatory instrument (CASANOVAS AND NORIEGA, 2022). As RUSSELL (2003) concludes, personal agents are owed to the users, not to the corporations that create them (*ibid.*: p. 71), but it may well be that in the end, if we do nothing to prevent it, we may end up consolidating "a great division between an elite assisted by humans, and a vast underclass, assisted, and controlled, by machines." (*ibid.*: p. 127).

What I think should be done is to rethink the conditions under which the legal technique can be activated, defining its implementation no longer as the application of law or norms, but directly as legal governance that operates from information systems. Rules and how they are constructed, emerge and operate are of utmost importance now. In digital environments, they are combined with the behavioural patterns that users build. We need to pay a closer attention to this relationship. Only in this way can the ideal of the rule of law cease to be an ideal and become the material regulation of the rights and duties of people in a world that, I am very much afraid, is not thought or designed for them.

IIIA-CSIC, Bellaterra, May 2024

**Acknowledgments**: I would like to thank, Lluís Payrató, Jordi Fortuny, and the *Grup d'Estudi de la Variació* (GEV) for extending this invitation to participate in the 31st Congress of General Linguistics of the University of Barcelona (CLUB31), held at UB Historical Building, on October 27[th], 2023. This work has been partially funded by the European Project, *Optimizing Manufacturing Processes through Artificial Intelligence and Virtualization*, OPTIMAI: Grant agreement ID: 958264 (https://optimai.eu/ ), and by the SGR 00532 and IIIA-UAB Associated Unit, *Institute of Law and Technology*. I thank Mustafa Hashmi for his invaluable cooperation and help, and my colleagues at IIIA-CSIC, especially Pablo Noriega, Enric Plaza, Ramon López de Mántaras, Lluis Godo and Carles Sierra, for the discussions held and the clarifications they have provided to me. Without them, this chapter would have contained much more ambiguities.